\documentclass[11pt]{article}
\usepackage[utf8]{inputenc}
\usepackage[T1]{fontenc}

\usepackage{amsfonts}
\usepackage{amsmath}
\usepackage{amssymb}
\usepackage{amsthm}
\usepackage{caption}
\usepackage{cite}
\usepackage{color}
    \definecolor{darkgreen}{rgb}{0,0.5,0}
    \definecolor{darkblue}{rgb}{0,0,0.6}
    \definecolor{purple}{rgb}{0.4,.2,0.7}
\usepackage[mathcal]{eucal}
\usepackage[margin = 2.5cm]{geometry}
\usepackage{graphicx}
\usepackage[hyperfootnotes = false, colorlinks = true, linkcolor = darkblue, citecolor = purple]{hyperref}
\usepackage{mathabx}
\usepackage{mathrsfs}
\usepackage{subcaption}

\renewcommand{\d}{\mathrm{d}}
\renewcommand{\i}{\mathrm{i}}

\begin{document}

\thispagestyle{empty}
\begin{center}
    ~\vspace{5mm}

    {\LARGE \bf {The double cone geometry is stable to brane nucleation}\\}

    \vspace{0.5in}

    {\bf Raghu~Mahajan,$^{1}$ Donald~Marolf,$^{2}$ and Jorge~E.~Santos.$^{3}$}

    \vspace{0.5in}

    $^1$ Department of Physics, Stanford University, Stanford, CA 94305, USA \vskip1em
    $^2$ Department of Physics, University of California at Santa Barbara, Santa Barbara, CA 93106,
USA \vskip1em
    $^3$Department of Applied Mathematics and Theoretical Physics, University of Cambridge, Wilberforce Road, Cambridge, CB3 0WA, UK

    \vspace{0.5in}

    {\tt raghumahajan@stanford.edu, marolf@ucsb.edu, jss55@cam.ac.uk}
\end{center}

\vspace{0.5in}

\begin{abstract}
\noindent
In gauge/gravity duality, the bulk double cone geometry has been argued to account for a key feature of the spectral form factor known as the ramp.
This feature is deeply associated with quantum chaos in the dual field theory.
The connection with the ramp has been demonstrated in detail for two-dimensional theories of bulk gravity, but it appears natural in higher dimensions as well.
In a general bulk theory the double cone might thus be expected to dominate the semiclassical bulk path integral for the boundary spectral form factor in the ramp regime. 
While other known spacetime wormholes have been shown to be unstable to brane nucleation when they dominate over known disconnected (factorizing) solutions, we argue that the double cone is stable to semiclassical brane nucleation at the probe-brane level in a variety of string- and M-theory settings.
Possible implications for the AdS/CFT factorization problem are briefly discussed.
\end{abstract}

\vspace{1in}

\pagebreak

\setcounter{tocdepth}{3}
{\hypersetup{linkcolor=black}\tableofcontents}

%%%%%%%%%%%%
\section{Introduction}\label{sec:intro}
%%%%%%%%%%%%

\begin{figure}[h]
\centering
\includegraphics[width =\textwidth]{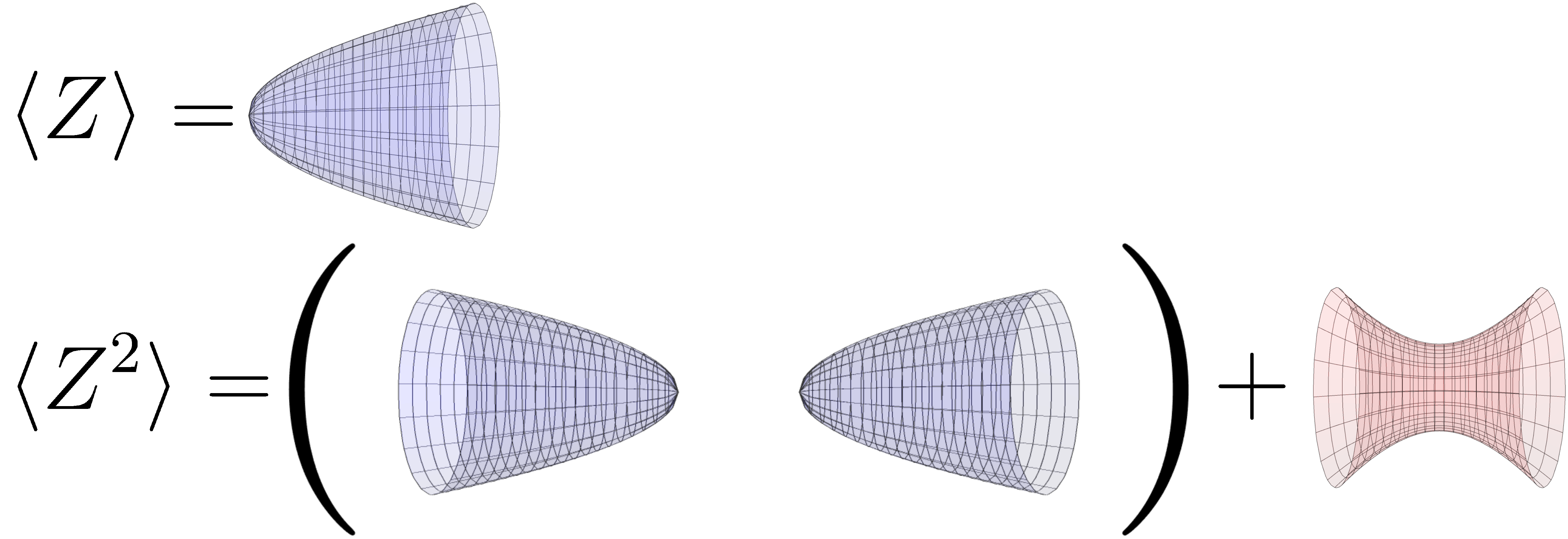}\\
\caption{An example showing failure  of factorization due to spacetime wormholes. The top line represents a path integral $\langle Z\rangle$. Although we have drawn the configuration as connected, it may include contributions from disconnected spacetimes as well. In any case, the natural path integral $\langle Z^2 \rangle$ associated with a pair of boundaries yields all terms generated by squaring $\langle Z\rangle$, but also contains additional contributions connecting the two boundaries as indicated by the second term in the bottom line.}
\label{fig:nofac}
\end{figure}

It has been understood since the late 1980's that contributions to gravitational path integrals from spacetime wormholes can markedly change properties of S-matrices, boundary correlators, or boundary partition functions \cite{Lavrelashvili:1987jg,Hawking:1987mz,Hawking:1988ae,Coleman:1988cy,Giddings:1988cx,Giddings:1988wv}.  
Here the term spacetime wormhole means a geometry whose boundaries have more than one connected component, and the discussion includes real geometries of any signature as well as spacetimes that are intrinsically complex. 
As a result, gravitational path integrals that do not include spacetime wormholes generally factorize into a product of terms associated with each boundary.   
But this factorization is expected to fail when spacetime wormholes contribute; see figure \ref{fig:nofac}.

One issue of interest is the potential consequence for the AdS/CFT correspondence.  
The standard picture of AdS/CFT in which bulk gravity is dual to a single well-defined CFT \cite{Maldacena:1997re,Gubser:1998bc,Witten:1998qj} would require the above factorization to hold, so failure of factorization requires a significant change.
In parallel with the understanding from the 1980's, a possible resolution is that the bulk gravitational path integral is dual to an ensemble of boundary theories; see e.g. recent discussions in
\cite{SSS-1,Saad:2019lba,Stanford:2019vob,Marolf:2020xie,Betzios:2020nry,Cotler:2020ugk,Maxfield:2020ale,Afkhami-Jeddi:2020ezh,Maloney:2020nni,Garcia-Garcia:2020ttf}.
For this reason, we use notation involving angle-brackets (e.g., $\langle Z \rangle$, $\langle Z^2 \rangle$, etc.) to denote results of bulk gravitational path integrals.   
A non-zero ``connected correlator'' $\langle Z^2 \rangle - \langle Z \rangle^2$ is then interpreted as describing $\delta Z^2$ for fluctuations $\delta Z$ associated with differences between the partition function $Z$ in any particular element of the ensemble and the ensemble-mean $\langle Z \rangle$.

While the importance of this issue has been understood for some time (see e.g. \cite{MaldacenaMaoz}), the issue has received renewed emphasis due to the recently-recognized role that replica wormholes play in allowing bulk gravitational path integrals to reproduce the Page curve associated with unitarity in black hole evaporation \cite{Almheiri:2019qdq,Penington:2019kki} and the associated connections to true spacetime wormholes and ensembles of boundary theories \cite{Penington:2019kki,Marolf:2020xie,Marolf:2020rpm}.  
The importance of understanding a possible ensemble interpretation of AdS/CFT was then further highlighted by the discovery \cite{Hsin:2020mfa,Chen:2020ojn} that it could circumvent the arguments of \cite{Harlow:2018jwu,Harlow:2018tng}.
In particular, while bulk theories dual to a single QM theory cannot have global symmetries, such symmetries {\it can} exist in bulk theories dual to an ensemble average.\footnote{
    Refs \cite{Hsin:2020mfa,Chen:2020ojn} describe senses in which the bulk theories nevertheless violate conservation of the global charge; see also \cite{Belin:2020jxr}.
    It is important to note that these violations arise when one focuses on charge in regions accessible to asymptotic observers.
    The full bulk state, including all so-called baby universes, appears to conserve global charge.
    We find this to be particularly clear from the real-time point-of-view described in \cite{Marolf:2020rpm,Colin-Ellerin:2020mva}.
}
This then raises the question of whether bulk theories dual to ensemble averages might also allow similar violations of other so-called swampland conjectures.

A critical question remains whether an ensemble interpretation will hold for the most familiar examples of AdS/CFT.  
The potential obstacle is that the relevant bulk theories involve large amounts of supersymmetry, and this supersymmetry should be reflected in each member of the boundary ensemble.\footnote{
    This property holds in the matrix model examples of \cite{Stanford:2019vob}.
    A general argument follows from the fact that the full bulk system admits an algebra of asymptotic SUSY charges, and that these charges must act trivially on the `baby universe sector' of the theory (i.e., on the ${\cal H}_{BU}$ of \cite{Marolf:2020xie,Marolf:2020rpm}).
    Thus the SUSY algebra acts within each bulk superselection sector. The boundary dual interpretation is then that each member of the associated ensemble has a well-defined SUSY algebra.
}
An ensemble interpretation is thus in tension with the idea that there are strict limits on SUSY CFTs in higher dimensions.  
In particular, for boundary dimension $d=4$ and ${\cal N}=4$ SUSY, there is a unique maximally-symmetric marginal deformation of appropriate free field theories which correctly predicts the derivatives of the $d=4$ ${\cal N}=4$ SU($N$) super Yang-Mills correlators evaluated at zero coupling.\footnote{
    This might indicate that  super Yang-Mills theory is the unique local maximally-supersymmetric theory that admits a weakly-coupled limit.   And such reasoning suggests to some that there should be a novel piece of physics associated with the UV-completion of bulk gravity that counteracts the effect of spacetime wormholes and allows the boundary ensemble to degenerate to a single unique theory; see related discussions in \cite{Marolf:2020xie,McNamara:2020uza}. Of course, such a resolution would be surprising from the viewpoint of low-energy semi-classical gravity. One alternative is that there is in fact an infinite discrete family of maximally supersymmetric CFTs with weakly coupled limits associated with different instanton corrections, perhaps due to a mechanism like that described by Seiberg in \cite{Seiberg:2010qd}.  We thank Juan Maldacena for correspondence on this point, and in particular for pointing out reference \cite{Seiberg:2010qd}.
}

It is natural to address this issue by investigating the status of bulk spacetime wormholes in UV-complete theories.  
In particular, it is interesting that analyses of Euclidean wormholes in 10- or 11-dimensional supergravity have always found such wormholes to fail to dominate path integral computations \cite{MaldacenaMaoz,Buchel:2004rr,Hertog:2018kbz,Betzios:2019rds,Marolf:2021kjc} (or else the analyses were unable to fully investigate the question).    
Two potential issues are that the wormhole can have `negative modes,' indicating that it fails to be a local minimum of the Euclidean path integral, and that adding various branes to the solution can lower the Euclidean action.
Whenever a Euclidean spacetime wormhole solution has been found to have lower action than the natural disconnected (factorizing) semi-classical saddle, investigations of such issues have found one or the other of these features to arise. 
However, it is also interesting that \cite{Marolf:2021kjc} identified sub-leading wormhole saddles that are free of both issues, and which thus appear to cause violations of factorization.  
The fact that such saddles are sub-leading would mean only that the associated violation is small, so that the relevant dual ensemble is sharply peaked.

It might be expected that any dual ensemble would be sharply peaked in this way. 
If this is the case, then bulk  wormholes should dominate only in the presence of very large sources that could compensate for the sharpness of this peak by spreading out the answers associated with various members of the ensemble.  
In general, such large sources could be associated with large back-reaction that is difficult to control.

However, as pointed out in \cite{SSS-1,Saad:2019lba,Stanford:2019vob}, studying the evolution of the system at large (Lorentzian) times can achieve a similar effect.  
Such large times may in fact be thought of as the large boundary sources suggested above, but in a context where back-reaction need not generate large curvature invariants. 
In particular, \cite{SSS-1,Saad:2019lba,Stanford:2019vob} identified a new saddle called the double cone which appears to dominate bulk path integral computations of the spectral form factor $\langle Z(\beta-iT) Z(\beta +iT) \rangle$ for large $T$, and which reproduces a crucial aspect (`the ramp') of the expected quantum-chaotic behavior of any dual theory associated with times $t_{\text{ramp}} \ll T \ll e^S$, where $t_{\text{ramp}}$ is non-universal and depends on the details of the system \cite{Gharibyan:2018jrp}.
Furthermore, due to its close relation (reviewed below) to AdS-Schwarzschild black holes, such double cone saddles will exist in any asymptotically-AdS bulk theory.

The double cone is thus a prime target for further investigation.
In particular, it is critical to understand the stability of such saddles from the bulk point of view.\footnote{
    Instability of the double cone would make wormholes less relevant, and would thus reduce the need for ensembles.
    But stability of the double cone would not forbid novel corrections from UV physics that could remove the need for an ensemble interpretation. One might imagine such corrections to leave the double cone as the dominant contribution to the average of the spectral form factor over some small window of time; see e.g. \cite{SaadTalk}.
    Furthermore, even without a possible link to ensembles, establishing stability of the double cone would make the bulk recovery of the ramp in \cite{SSS-1,Saad:2019lba,Stanford:2019vob} more robust.
}
The double cone was not included in the analysis of \cite{Marolf:2021kjc} due to various subtle features and the fact that it is naturally viewed as having a complex metric.
We take up the challenge below and argue at the probe-brane level in various UV-complete contexts that the double cone does {\it not} suffer from brane-nucleation instabilities for $S^{d-1} \times S^1$ boundary metrics.\footnote{
    In contrast, branes give a divergent contribution to the double cone amplitude with flat boundaries.  This is consistent with the boundary expectation that there is no ramp in such cases due to a divergence in the density of states associated with the infinite volume of a non-compact moduli space.  This divergence also leads to a continuous spectrum.
}

However, we do not address the question of possible field-theoretic negative modes, as this is especially complicated to understand in complex spacetimes; see further discussion in section \ref{sec:over}. 
Another analysis of double cone stability based on a constrained-instanton description is also being released simultaneously with this work \cite{CJnew}.

We begin our discussion with a conceptual overview in section \ref{sec:over}.
The emphasis here is on the double cone geometry described as a complex saddle for the bulk path integral that computes the boundary spectral form factor, and on understanding what it means to study brane-nucleation stability issues on complex spacetimes.  
We in particular mention subtleties associated with brane back-reaction and the microcanonical ensembles used to define the spectral form factor, though we defer consideration of back-reaction to section \ref{sec:disc} and focus on strict probe-branes in the bulk of this work.
The rest of the stage for our calculations is set by establishing conventions and reviewing brane actions in section \ref{sec:coven}.  
We then proceed to study the AdS$_5$ double cone ($\times S^5$) as a saddle for 10d type IIB supergravity in section \ref{sec:ads5}.  
In this context, the branes whose nucleation might seem most likely to lead to an instability are supersymmetric D3-branes.  
But at the probe-brane level we argue against such instabilities by showing that all saddle-points of the D3 brane action (in the full complexified double cone geometry) have positive real part for the relevant Euclidean action.
We also use this section to review the general construction of the double cone spacetime and the sense in which it is intrinsically complex.  
Section \ref{sec:gend} then studies generalizations to other truncations of 10- and 11-dimensional supergravity that allow AdS$_{d+1}$-Schwarzschild double cones, and section \ref{sec:CMP} further generalizes to charged (boundary dimensions $d=3$ and $d=4$) and rotating (boundary dimension $d=4$) double cones. 
Similar cases with flat boundaries (and in particular BTZ double cones) are discussed in appendix \ref{app:btz}. 
We close with a discussion of back reaction and other open issues in section \ref{sec:disc}.

%%%%%%%%%%%%
\section{Overview}
\label{sec:over}
%%%%%%%%%%%%

This section provides a conceptual orientation to the physics that will be studied in the rest of this work.
We first review the double cone geometry in its role as an intrinsically-complex saddle for the bulk path integral computing the so-called spectral form factor.
We then describe what it will mean to study the stability of such complex saddles with respect to brane nucleation.

\subsection{The spectral form factor and the double cone}

The spectral form factor (SFF) for any quantum-mechanical theory is defined by studying the partition function
$Z(\beta) = {\rm Tr} \left(e^{-\beta H} \right)$
at imaginary $\beta =-\i T$, or more generally at complex $\beta$.
In particular, the SFF is the product
$Z(-\i T)Z(\i T) = {\rm Tr}  \left(e^{-\i HT} \right){\rm Tr} \left(e^{\i HT} \right)$.
Furthermore, it is often useful to restrict the trace to states with energies in some window $[E_1,E_2]$, or equivalently to evaluate the trace in a microcanonical ensemble.
This in particular guarantees the traces to converge in any theory with a finite density of states.

Refs \cite{SSS-1,Saad:2019lba,Stanford:2019vob} study the bulk gravitational path integral that would compute the SFF in a dual theory (and with any baby universe sector in its Hartle-Hawking state in the language of \cite{Marolf:2020xie}).
Using the angle-bracket notation above, they thus study $\langle Z(-\i T)Z(\i T) \rangle= \langle {\rm Tr} \left(e^{-\i HT} \right){\rm Tr} \left(e^{\i HT} \right)\rangle$. 
In particular, the above references consider bulk gravitational systems with two Lorentz-signature asymptotically AdS boundaries, each of the form $S^1 \times X$.  The $S^1$ factor  is timelike and is periodic with boundary time $T$, but with time running in opposite directions around the $S^1$ factors on opposite boundaries.
The latter requirement is perhaps best understood as requiring the bulk path integral to have a symmetry that complex conjugates the bulk amplitude and exchanges the two boundaries.
As suggested above, microcanonical boundary conditions are also imposed, forcing the total energy at each boundary to lie in a small window near some value $E$.

\begin{figure}[htb]
    \centering
    \includegraphics[width=0.8
\textwidth]{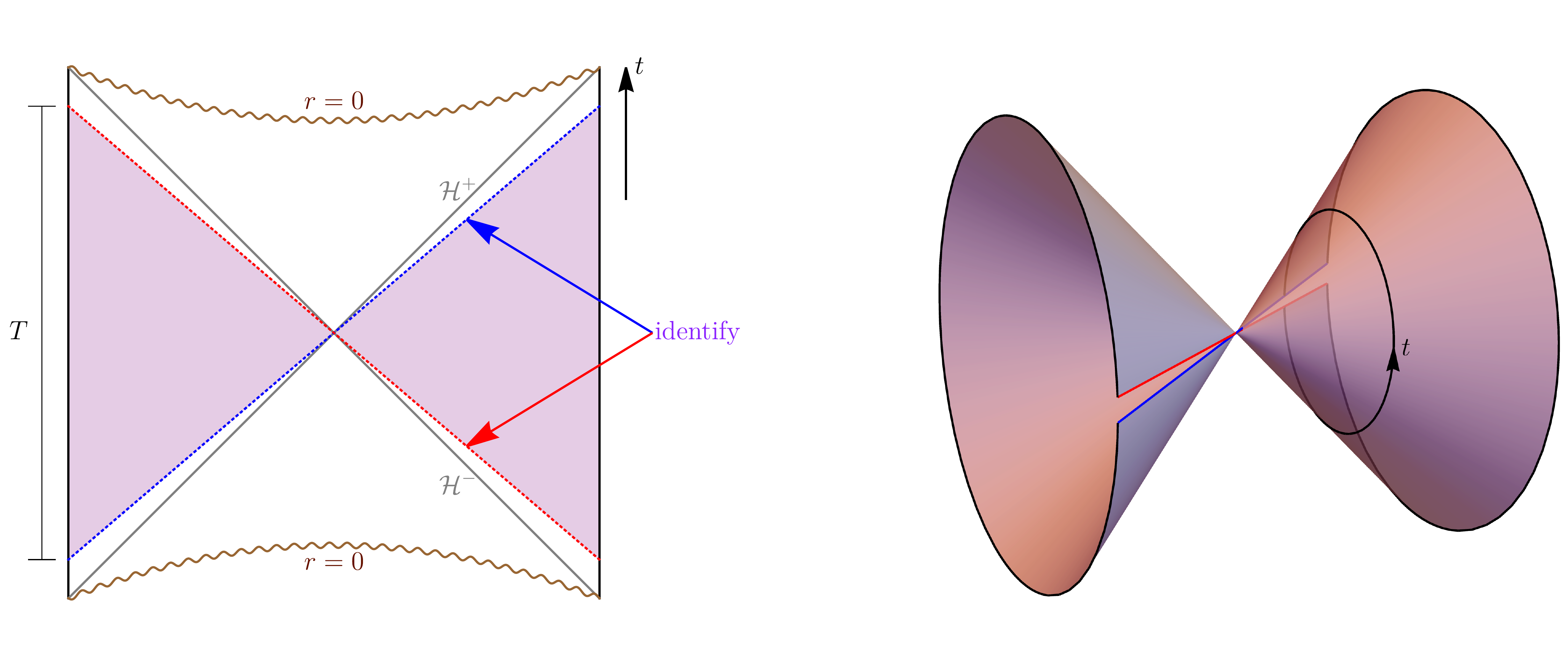}
    \caption{Schematic illustration showing the identifications in the Kruskal spacetime that lead to the double cone geometry.}
    \label{fig:ident}
\end{figure}

The above references then consider any two-sided asymptotically AdS stationary black hole (with a bifurcate Killing horizon); e.g., perhaps the Kruskal-like extension of AdS-Schwarzschild.  
They note that periodically identifying such a solution under translations of magnitude $T$ generated by the stationary Killing field $\partial_t$  yields a spacetime that satisfies the desired boundary conditions.  
In particular, because $\partial_t$ is future-directed in one asymptotic region and past-directed in the other, the quotient naturally has $S^1$ factors whose periods correspond to boundary times $T$ and $-T$ as desired; see figure \ref{fig:ident}.

A complication, however, is that $\partial_t$ has fixed points on the bifurcation surface.  
This makes the quotient singular, raising potential questions about its validity as a saddle point in the path integral.  
In particular, it is not immediately clear how to check whether the action is in fact stationary under variations of the metric at this singularity. 
But reference \cite{SSS-1} regularizes the singularity by making an excursion into the complex plane.  
Since the fixed points of $\partial_t$ in the full complexified black hole geometry all lie on the bifurcation surface of the real Lorentzian slice, any contour that avoids this surface (and which also avoids the black hole singularity) and which also remains invariant under $t \rightarrow t+T$ yields a smooth quotient. 
Indeed, so long as they also end on the asymptotically AdS boundary of the original real slice, quotients of such contours define (complex) solutions which satisfy all boundary conditions imposed by the path integral.  
Thus any such contour defines a valid complex saddle for the SFF path integral.

\begin{figure}[htb]
    \centering
    \includegraphics[width=0.3\textwidth]{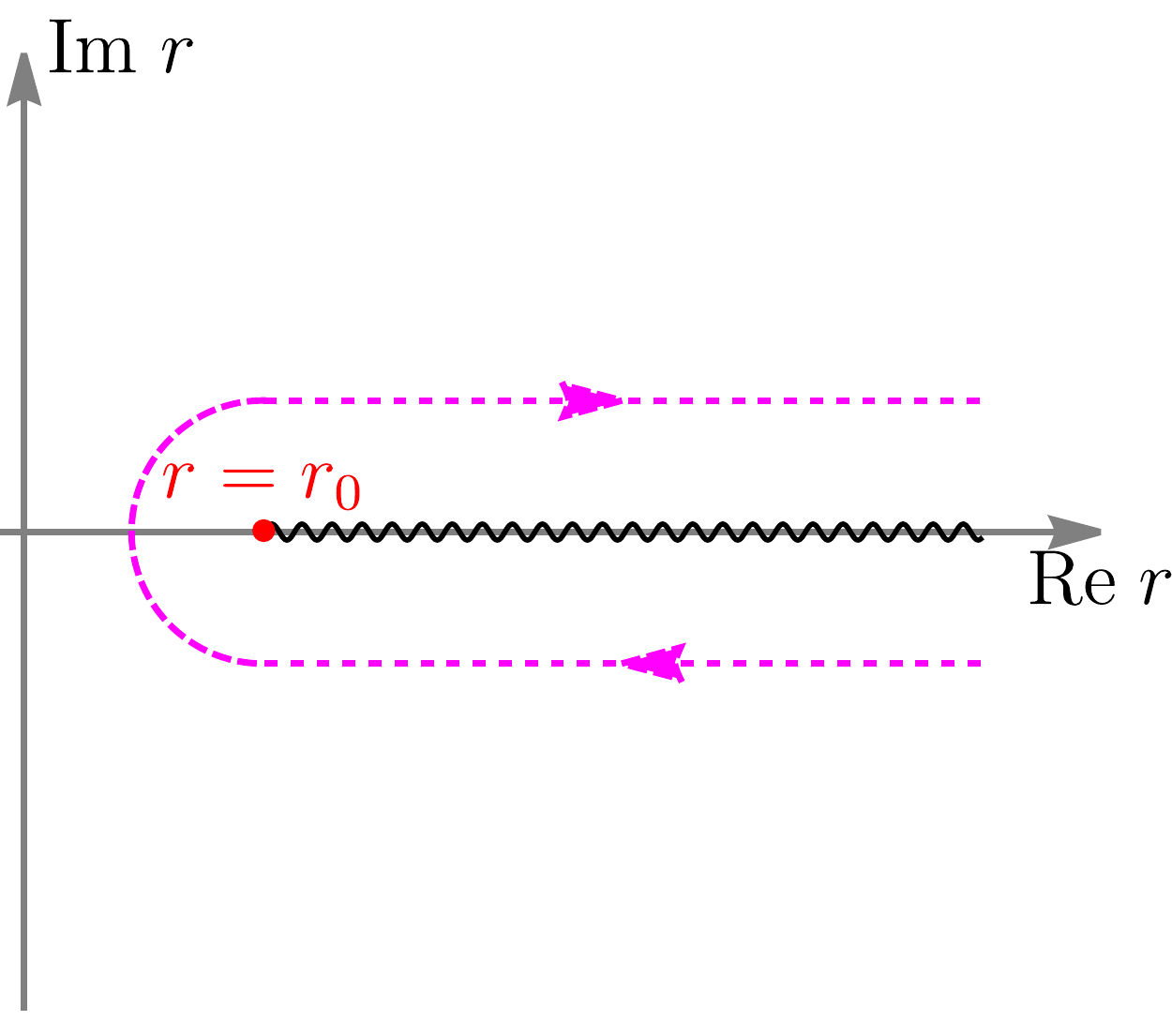}
    \caption{A keyhole-type contour in the complex $r$ plane that defines the double cone.}
    \label{fig:sub1}
\end{figure}

As described in \cite{SSS-1}, finding such contours is not difficult.  
In particular, for any two-sided black hole one may consider the Schwarzschild area-radius $r$, noting that two distinct surfaces in the real black hole spacetime map to each value of $r$ so that the map from $r$-values to the spacetime has a branch point at some $r=r_0$ describing the bifurcation surface.  
One may then consider a contour in the complex $r$ plane that starts at real positive infinity and approaches close to the (real) value $r_0$, but which then circles halfway around $r_0$ through the complex plane and returns to positive real infinity (now in the opposite asymptotic region due to the branch point at $r_0$, see figure \ref{fig:sub1}).
The geometry on this contour fulfills all of the conditions required above.

We will thus use the term ``double cone'' to refer to any such smooth complex contour in the periodically identified complexification of a two-sided stationary black hole.  
In this way, it is natural to think of the double cone as a spacetime with an intrinsically complex metric.  
Note that, while there is great ambiguity in the choice of this contour, Cauchy's theorem guarantees the action to be invariant under deformations of the contour.  
As a result, this ambiguity does not affect the result of our path integral.  
The action for simple cases is computed in \cite{SSS-1,Saad:2019lba,Stanford:2019vob}, and is argued to have the right properties to dominate the semiclassical bulk computation of the spectral form factor in the ramp regime $t_{\text{ramp}} \ll T \ll e^S$ \cite{Gharibyan:2018jrp}.

\subsection{Brane nucleation on complex saddles}
\label{sec:saddlesanddescents}

Although the double cone is indeed a saddle, and while it reproduces the expected properties of the SFF ramp, it is important to investigate whether the double cone is in fact the dominant semiclassical bulk saddle in the ramp regime, or whether some other saddle or off-shell effect might provide a larger contribution that would make the double cone irrelevant.
For a real saddle in a Euclidean path integral, one can probe this by investigating whether small real perturbations of the saddle increase or decrease the action.\footnote{
    One should also take proper care of issues \cite{Gibbons:1976ue,Gibbons:1978ji,Hartle:1988xv,Gratton:1999ya,Khvedelidze:2000cp,Gratton:2000fj,Gratton:2001gw,Dasgupta:2001ue,Anninos:2012ft,Cotler:2019nbi,Benjamin:2020mfz} related to the so-called conformal factor problem which makes the action for Euclidean gravity unbounded below.
    See also \cite{Kol:2006ga,Monteiro:2008wr} for discussions of such issues in the presence of couplings to matter.
}
In this context, small real perturbations of the bulk supergravity fields that decrease the Euclidean action are known as (field-theoretic) negative modes.

On the other hand, as is familiar from the general properties of analytic functions, near any complex saddle there are necessarily certain complex perturbations that decrease the real part of the Euclidean action as well as those that increase the real part.
One could thus say that all saddles have negative modes in the complex plane, but this is clearly neither useful nor physical relevant.
Indeed, the same is true of the above real Euclidean saddles if one considers both the original real perturbations and the imaginary perturbations obtained by multiplying them by $\i$.

Instead, for a complex saddle the first relevant question is whether it lies on what we will call a `global descent contour,' by which we simply mean a contour along which (1) the integrand has constant phase (so that the imaginary part of the Euclidean action is constant), (2) the saddle in question maximizes the integrand (and thus minimizes the real part of the Euclidean action), and (3) the real part of the Euclidean action becomes infinitely positive at a sufficient rate as one follows the contour to an appropriate notion of infinity.
If so, then the saddle will dominate the integral along the global descent contour in the semiclassical limit  $\hbar \rightarrow 0$ (here $\ell_{\text{Planck}} \rightarrow 0$).

Finally then, one should ask if the global descent contour can be obtained by deforming the defining contour of the integral without passing through singularities or picking up contributions from `arcs at infinity' or other boundaries of the contour.
If this can be done, then Cauchy's theorem tells us that our complex saddle can be used to compute a good semiclassical approximation to the desired integral.
Unfortunately, for an infinite-dimensional path integral such issues are clearly extremely complicated to analyze in full, and are beyond the scope of this work.
We thus set such concerns aside.

However, another issue turns out to be more important for most known examples of real Euclidean spacetime wormholes in UV-complete theories.
In the context of string theory or M-theory, one can imagine starting with the original saddle and building a new spacetime by adding various D-, NS-, or M-theory branes.
It is useful to stress that, at least as the limit of smooth regulated solutions, one can replace the branes by appropriate bulk fluxes (in the sense of \cite{Polchinski:1995mt}, see also related discussions in \cite{Kaplan:1995cp,Simon:2011rw}).  
In particular, we can think of this construction as producing a new bulk supergravity spacetime with a different topology (and with the above fluxes threading a new `handle').
At the conceptual level, the main difference from the above discussion of negative modes is then just that negative modes involve infinitesimal versions of  deformations of the original spacetime, while the addition of branes requires a discrete change in topology and/or quantized charges.

This difference has several important consequences.
The first is that, in contrast to the field-theoretic perturbations discussed above, one cannot simply multiply a brane-induced deformation by $\i=\sqrt{-1}$.
In particular, even in a complex spacetime we need consider only branes that have real tension and charge.
The reality (and in fact integer-valuedness) of the charge is especially manifest in a magnetic description where it is both real and quantized due to topological features, and in string- or M-theory the ratio of the tension to (the absolute value of) the charge is just a given (real) function of the (real) coupling constants of the theory.
So the addition of any particular configuration of branes will have an unambiguous effect on the Euclidean action.

The second consequence of the discreteness of branes is that, even if we start with a bulk saddle, at weak coupling the change in the action under the addition of branes is generally of first order in the brane charge $Q$.
Furthermore, because continuous deformations of a saddle can affect the action only at second order, this first-order-in-$Q$ change is independent of any first-order-in-$Q$ modifications one chooses to make in the original saddle in the region far from the new brane sources (or far from the associated changes in topology).
As a result, the first-order-in-$Q$ change in the bulk action under the addition of branes is a local function of the brane configuration to be added.
Indeed, in a weakly-coupled limit this change must agree with the corresponding probe-brane action evaluated on the desired configuration; see again \cite{Polchinski:1995mt,Kaplan:1995cp,Simon:2011rw}).
The relevant probe-brane actions will be reviewed in section \ref{sec:coven} below.

Note, however, that this probe-brane action will generally be an analytic function of parameters describing the configuration of branes, and in particular of the location at which the branes are inserted.
And as in our discussion of field-theoretic perturbations, the real part of an analytic function cannot be bounded either above or below, making it clear that the real part of the Euclidean probe brane action will become arbitrarily negative for branes inserted at appropriate complex positions in a complexified spacetime.
But once again, this issue cannot be of physical relevance.

Indeed, the key point here is that we have so far used the insertion of branes to create only new off-shell complex configurations for our path-integral.
What we wish to understand is the integral of the associated amplitude over an appropriate contour.
At the semiclassical level, this as always becomes the questions of what saddle points might exist, whether they lie on global descent contours, and whether such global descent contours can be obtained by allowed deformations of the original contour of integration.

We thus focus below on identifying saddles of the probe brane action on the complexified spacetime.\footnote{
    In studying brane nucleation in Euclidean signature one often does not bother to identify saddles, but instead declares an instability when any configuration has negative (real) Euclidean action.
    This is justified since any configuration with Euclidean action requires there to be a local minimum (defining a saddle) or divergent contributions from asymptotic regions of the path integral with negative action.
}
We assume such saddles to be invariant under the Killing symmetries of the double cone (or, at least those associated with the AdS$_{d+1}$  factor of the spacetime).
This then reduces the space of configurations so that it can be parametrized by a single complex coordinate (e.g., the Schwarzschild area-radius $r$).
We can then identify global descent contours within this one-dimensional complex space, and we can analyze whether such contours can be deformed to the contour we use to define the double cone (or, equivalently, whether it can be deformed to the positive real $r$-axis; see figure \ref{fig:sub1} above or the pink dashed lines in figures \ref{fig:ads4} and \ref{fig:subcharged} below).

An important point is that, whenever we find a global descent contour that can be deformed to the defining contour, Cauchy's theorem guarantees that the integral along this contour gives the desired answer. 
As a result, in the semiclassical limit our path integral {\it will} be dominated by the relevant saddle.  
In particular, if this saddle has positive Euclidean action then the result of the integral will be small in the semiclassical limit.  
It therefore follows that no saddle with negative Euclidean action can dominate.  
Instead, any global descent contours associated with saddles having negative Euclidean action must fail to be deformable to the defining contour.  
It is thus not necessary to find all saddles when one can study such contours in detail.
After setting conventions and reviewing certain technical details in section \ref{sec:coven}, we execute this procedure for the AdS$_5\times S^5$ double cone in section \ref{sec:ads5} and for other string- and M-theoretic cases in section \ref{sec:gend}.

\subsection{Branes and anti-branes}
\label{sec:bab}

We now pause to explain an important further subtlety in our analysis.
With boundary conditions that fix only potentials at infinity, solutions typically exist in which one adds only a single brane and thus in which one changes the asymptotic charges on one side or the other.
Recall, however, that we wish to impose microcanonical boundary conditions which fix the total energy on both sides of the black hole.  
We also wish to fix the D-brane charge on each boundary, corresponding to fixing at least simple parameters describing the theory in which the SFF is computed (e.g. N for an SU(N) gauge theory).  
At least on-shell, this is inconsistent with the addition of a single brane when one goes beyond the strict probe-brane approximation to allow the new branes to contribute to the total bulk charges.

Now, we can easily arrange to satisfy the D-brane charge constraint by adding instead a brane/anti-brane pair for which the total D-brane charge vanishes.  
However, the energy constraint turns out to be harder to satisfy in a useful way.  
It is true that on the real section the background has a CPT symmetry that swaps branes for anti-branes and interchanges the two boundaries.  
So for a real brane saddle at area-radius $r_\star$ with energy $E_\star$, there is a corresponding anti-brane saddle on the `second sheet' at $r_\star$ with energy $-E$ (as defined with respect to the Killing field $\partial_t$).  
And by analytic extension this remains true of complex saddles.  
However, the same symmetry implies the probe-brane/anti-brane actions to be $I$ and $-I$.  
Thus the pair contributes a net zero action which one might think suggests an instability.  
But since this symmetry is also valid off-shell, the change in sign of the action means that it maps descent contours for the brane to {\it ascent} contours for the anti-brane and vice versa.  
This in turn suggests that if e.g. the brane descent contour can be deformed to the defining contour then this will not be possible for the anti-brane descent contour.

Indeed, a sharp argument can be made by noting that the Lorentz-signature probe-brane/anti-brane actions are real on the real section.  
They thus extend to functions on the complex $r$-plane that satisfy $I_L(r^*) = \left(I_L(r)\right)^*$, with $^*$ denoting complex conjugation.
For the Euclidean action $I_E = -\i I_L$, this implies $I_E(r^*) = - \left(I_L(r)\right)^*$.  
So if the brane saddle at $r_\star$ has actions $I_{L\star},I_{E\star}$, then the anti-brane saddle on the 2nd sheet at $r_\star^*$ has actions $-I^*_{L\star},I^*_{E\star}$.  
Thus for this pair of brane/anti-brane saddles our symmetry preserves the magnitude of the path integral amplitude and so maps descent contours to descent contours.  
Furthermore, since the defining contour for the double cone is invariant under the CPT-like map defined by including this extra complex conjugation, the brane descent contour through $r_\star$ can be deformed to the defining contour if and only if the same is true for the anti-brane descent contour through $r_\star^*$ on the second sheet.  
In other words, using the argument from the subsection above, it is guaranteed that any pair of brane/anti-brane saddles that dominate our path integral will be of this form.  
In particular, if the contribution from this latter pair is suppressed, then the zero-action pair described above that both lie at $r_\star$ cannot contribute.

We will thus study brane/anti-brane pairs of this latter form (brane at $r_\star$, anti-brane on the second sheet at $r_\star^*$) in all cases below. 
And as foreshadowed above, we will always find the saddles with useful descent contours to have ${\rm Im} I_L >0$ (and so also ${\rm Re} I_E >0$) so that these contributions are indeed suppressed.  
At the level of a strict probe-brane calculation this then indicates stability of the double cone.  

However, when one considers back-reaction one must return to the issue of the microcanonical boundary conditions.  
Under the various symmetries above, the energy and action transform in precisely the same way.  
This is to be expected, as for the static solutions we consider the brane action takes the form $I_L =\int \left( p\dot{q}-H_{\text{brane}} \right) = - H_{\text{brane}}T$, with $H_{\text{brane}} = E +q\Phi$ in terms of the energy $E$, the brane-charge $q$, and the corresponding potential $\Phi$.  
The upshot is then that if our brane at $r_\star$ has energy $E_\star$, then the  anti-brane at $r_\star^*$ has energy $-E_\star^*$ with respect to the KVF $\partial_t$.  The total energy contributed by the pair is thus $E_\star -E_\star^* = 2\i \, {\rm Im} E_\star$, which will not generally vanish.

In many perturbative problems one could compensate for any non-zero boost energy contributed by the branes by adjusting parameters in the double cone background.  
However, on-shell double cones exist only when the boost energy vanishes exactly.  
So shifting to some non-zero (imaginary) value is not allowed.  
The physics of this result is essentially the same as that for the brane charge, where Gauss' law requires a strict source-free on-shell double cone to have equal and opposite charges at the right and left boundaries.

As a result, unless $E_\star$ happens to be real, it will not be possible to incorporate perturbative back-reaction from our branes using on-shell techniques.  
The correct approach should be to include off-shell physics, but that is beyond the scope of this work.

In sections \ref{sec:ads5}-\ref{sec:CMP} we thus confine ourselves to a strict probe-brane analysis without considering brane contributions to the total charges or their effect on the microcanonical boundary conditions.
The fact that we find all relevant saddles to be suppressed even in this less stringent context is a strong argument for stability of the double cone. 
We will, however, briefly return to the issue of back-reaction in section \ref{sec:disc}, where we highlight the off-shell information required for a full analysis (though we leave its investigation for future work).

\subsection{Remarks about boundary curvature.}
We conclude this already-lengthy overview section with a short remark about the role of boundary curvature in brane nucleation instabilities in AdS.
As pointed out in \cite{MaldacenaMaoz} (see their equation (3.7)), if the asymptotically AdS space has a negative curvature boundary, then the brane action is such that moving the branes closer to the boundary lowers the Euclidean action.
The branes thus want to fly off to infinity.
This instability is reflected in the boundary theory via conformal couplings of scalar fields, which act as negative mass-squared terms if the boundary curvature is negative; see e.g. \cite{Maldacena:1998uz,Seiberg:1999xz,Maldacena:2000hw}.

The story is different when the boundary curvature is positive.
From the point of view of the brane action, equation (3.7) of \cite{MaldacenaMaoz} is then modified by the replacement of $\cosh \rho$ by $\sinh \rho$:
\begin{align}
    (\sinh \rho)^d - d \int^\rho \d \rho' (\sinh \rho')^d \sim + \frac{2d}{2^d(d-2)} \,
    e^{(d-2)\rho} \, ,
\end{align}
where crucially the sign is now different than in \cite{MaldacenaMaoz}.
This means that the branes do not want to run away to the boundary, and this agrees with the conformal couplings of the scalars in the boundary CFT now endowing them with a positive mass-squared.
Of course, the above provides only an asymptotic analysis near the AdS boundary.
The goal of this paper is to perform a systematic analysis of the saddle points of the brane action everywhere on the double cone geometry with positive curvature boundaries.
In this context we find no brane nucleation instabilities.
Double cones with flat boundaries are treated in the appendix.

%%%%%%%%%%%%
\section{Conventions and Actions \label{sec:coven}}
%%%%%%%%%%%%

This work studies configurations of D$p$-branes and M-branes in various models. 
The two cases are very similar, and we discuss each in turn.

The D$p$-brane configurations descend from the supersymmetric D$p$-brane action
\begin{equation}
I_{\mathrm{D}p} = - \int \mathrm{d}^{p+1}\,\sigma \,e^{-\frac{p-3}{4}\phi}\,\sqrt{-\mathrm{det}\,\mathcal{G}}+\int \mathcal{C}_{(p+1)},
\label{eq:susyborn}
\end{equation}
where $\mathcal{G}$ is the pull-back of the target space metric in the Einstein frame to the brane world-volume, $\phi$ is the dilaton and
\begin{equation}
\mathcal{C}_{(p+1)}=\frac{1}{(p+1)!}\varepsilon^{\mu_1\mu_2\ldots \mu_{p}\mu_{p+1}}\frac{\partial x^{a_1}}{\partial \sigma^{\mu_1}}\frac{\partial x^{a_2}}{\partial \sigma^{\mu_2}}\ldots \frac{\partial x^{a_p}}{\partial \sigma^{\mu_p}}\frac{\partial x^{a_{p+1}}}{\partial \sigma^{\mu_{p+1}}}C_{(p+1)\;a_1 a_2\ldots a_p a_{p+1}},
\end{equation}
where $C_{(p+1)}$ is the Ramond-Ramond (RR) target space potential.  
In particular, we can consistently ignore contributions of the gauge field on the brane to the Born-Infeld term as it will not be sourced by the background in the configurations studied below.  
The action for anti-branes differs only by a sign in the second term.
Below, we focus on backgrounds with a $\mathbb{Z}_2$ symmetry exchanges the two boundaries and which also exchanges branes and anti-branes.  
As a result, as described in section \ref{sec:bab}, it suffices to ignore anti-branes and to compute the action for branes with the above signs so long as we allow such branes to lie on either side of our double cones.  For simplicity we have set the brane tension to unity in \eqref{eq:susyborn}.
We have written the action in Lorentzian signature, and we note that the path integral weighting is
\begin{equation}
\exp{\i\,I_{\mathrm{D}p}}\,.
\end{equation}
The corresponding Euclidean action is thus $I^{\text{Euclidean}}_{\mathrm{D}p}= -\i I_{\mathrm{D}p}$.

In sections \ref{sec:ads5} and \ref{sec:gend}, we will be interested in adding such branes to spacetimes with geometry SAdS$_{d+1}\times X$, where SAdS$_{d+1}$ is a $(d+1)$-dimensional AdS-Schwarschild spacetime and $X$ is a compact manifold.  
In particular, with spherical boundaries the metric on SAdS$_{d+1}$ takes the form
\begin{equation}
\mathrm{d}s^2=-f(r)\mathrm{d}t^2+\frac{\mathrm{d}r^2}{f(r)}+r^2\mathrm{d}\Omega_{d-1}^2\,,
\label{eq:refanyd}
\end{equation}
with $\mathrm{d}\Omega^2_{d-1}$ is the unit round metric on a $(d-1)$-dimensional sphere and
\begin{align}
    f = 1 + \frac{r^2}{L^2} - \frac{2M}{r^{d-2}} \, , \quad  2M = r_+^{d-2}\left(1+\frac{r_+^2}{L^2}\right)\, ,
\end{align}
in terms of the horizon area-radius $r_+$ and the AdS scale $L$.
The full solution will also involve various matter fields so as to solve the relevant supergravity equations of motion.  
In particular, AdS compactifications of 10-dimensional supergravity are associated with fluxes for a U(1) gauge field that -- in an electric representation -- may be thought of as a $d$-form potential with $(d+1)$-form field strength $F$ supported on the SAdS$_{d+1}$ factor.\footnote{
    In some cases there may be additional fluxes as well, though they are typically related to the aforementioned flux by some sort of symmetry (at least after Kaluza-Klein reduction on $X$). 
}
For any given model, there are a variety of branes that we could consider adding in various configurations.
However, one expects that the branes most likely to activate an instability are those where the above flux acts to lessen the branes' attraction to the black hole.
We thus focus below on branes with $p=d-1$ with the $d = (p+1)$ spacetime dimensions along the brane oriented to lie ``in the same directions'' as the conformal boundary of the SAdS$_{d+1}$ factor.
We also take the brane configuration to preserve the symmetries of the SAdS$_{d+1}$ factor, so in practice this places each brane at a constant value of the Schwarzschild area-radius coordinate $r$.

The above features are common to each model below.  
Furthermore, we restrict attention to models with vanishing dilaton ($\phi=0$) and which live in supersymmetric theories.  
The latter condition fixes the relative normalization of the two integrands in \ref{eq:susyborn} at the conformal boundary of SAdS$_{d+1}$.  
Taken together, these features allow a uniform treatment of all string-theoretic models.

As an illustrative example, we now supply further details for the familiar case $p=3$ in which we consider type IIB supergravity solutions that are asymptotically AdS$_5 \times S^5$.
In the relevant solutions the only non-trivial bosonic fields are the self dual five-form $F_{(5)}$ and the ten-dimensional metric $G$. 
For this particular sector, the equations of motion read
\begin{subequations}
\begin{align}
&F_{(5)}=\mathrm{d}C_{(4)}
\\
&F_{(5)}=\star_{10} F_{(5)}
\\
& {}^{G}R_{AB}=\frac{1}{96}F_{(5)\;ACDEF}F_{(5)\;B}^{\phantom{(5)B}CDEF}
\end{align}
\end{subequations}%
where $\star_{10}$ is the ten-dimensional Hodge operation with respect to $G$, ${}^{G}R_{AB}$ are the components of the Ricci tensor associated with $G$, $C_{(4)}$ the RR 4-form potential and upper case Latin indices are ten-dimensional.

For most of this work we are interested in studying configurations arising in five-dimensional minimal gauged supergravity, whose action comprises a five-dimensional metric $g$ and a five-dimensional gauge field $F=\mathrm{d}A$
\begin{equation}
I_{5D}=\frac{1}{16 \pi G_5}\int_{\mathcal{M}}\sqrt{-g}\left(R+\frac{12}{L^2}-\frac{1}{4}F^{ab}F_{ab}+\frac{1}{12\sqrt{3}}\varepsilon^{abcde}F_{ab}F_{cd}A_e\right)\,,
\label{eq:1}
\end{equation}
where $R$ is the Ricci scalar associated with $g$, lower case Latin indices are five-dimensional indices and $L$ is the five-dimensional cosmological constant.

Solutions to the equations of motion derived from Eq.~(\ref{eq:1}) are known to uplift to solutions of type IIB supergravity via \cite{Chamblin:1999tk,Cvetic:1999xp,Cvetic:2000nc}
\begin{subequations}
\begin{align}
&\mathrm{d}s^2= g_{ab}\,\mathrm{d}x^a\,\mathrm{d}x^b+L^2\left[\left(\mathrm{d}\Psi+\mathbb{A}-\frac{A}{\sqrt{3}L}\right)^2+\mathrm{d}S_{\mathbb{CP}_2}^2\right]
\\
&G_{(5)}=-\frac{4}{L}\mathrm{Vol}_{5}-\frac{L^3}{2\sqrt{3}}\mathbb{J}\wedge \star_5 F\,,
\\
&F_{(5)}=G_{(5)}+\star_{10} G_{(5)}\,,
\end{align}
\label{eq:consi}%
\end{subequations}
where $\mathrm{Vol}_5$ is the volume form of $g$, $\star_5$ is the five-dimensional Hodge dual operation obtained using $g$, $\mathrm{d}S^2_{\mathbb{CP}_2}$ is the standard Fubini-Study metric on $\mathbb{CP}^2$ and $\mathbb{J} = \mathrm{d}\mathbb{A}$ is its associated K\"ahler form. 
Note that if $A=0$ one recovers the standard Freund–Rubin compactification \cite{Freund:1980xh}. 
One can then use the above to compute the action \ref{eq:susyborn} for D3 branes at constant $r$ and constant location on the $S^5$ to find that the brane contributes a path integral weight
\begin{align}
    \exp \i I_{{\mathrm D}p}  = \exp \left\{
    - \i\,T\,\Omega_{d-1}\left[r^{d-1}\sqrt{f(r)} - \frac{r^d - r_+^d}{L}  \right]
    \right\},
    \label{eq:alld}
\end{align}
where $\Omega_{d-1}$ is the area of a unit round $(d-1)$-dimensional sphere and we have written the answer in a form where it in fact applies to all models considered below for any $p=d-1$.
Again we remind the reader that we have set the brane tension to unity.

In fact, the expression \eqref{eq:alld} also applies to M-theoretic models.
Since M-theory contains M$2$-branes and M$5$-branes, such models are relevant to $d=3$ and $d=6$.  
In particular they pertain to configurations that result from embedding Schwarzschild-AdS black holes with $d=3$ and $d=5$ in eleven-dimensional supergravity. 
From the higher-dimensional perspective, these backgrounds asymptote to AdS$_4\times S^7$ and AdS$_7\times S^4$, respectively. 
In that case the supersymmetric probe-brane action is \begin{equation}
\tilde{I}_{\mathrm{M}p}=-\int \mathrm{d}^{p+1}\,\sigma\,\sqrt{-\mathrm{det}\,\mathcal{G}}+\,\int \mathcal{C}_{(p+1)}\, ,
\label{eq:dbranefinalmem}
\end{equation}
with $p=2$, $p=5$ for the AdS$_4\times S^7$ and AdS$_7\times S^4$ cases, respectively.
Eq. \eqref{eq:dbranefinalmem} differs from \eqref{eq:susyborn} mostly by deleting the dilaton factor. 
However, since we restrict analysis below to those string-theoretic cases in which the dilaton vanishes, our treatment of M-branes will be essentially identical to our treatment of D-branes.

In particular, it turns out that Eq.~(\ref{eq:alld}) again describes the weight that each M-brane contributes to the path integral.  
This can be verified in detail using the standard Freund–Rubin compactification \cite{Freund:1980xh} to relate the AdS$_4$ and AdS$_7$ theories to eleven-dimensional supergravity.  
For reference purposes, our conventions are that the equations of motion of eleven-dimensional supergravity read
\begin{subequations}
\begin{align}
& {}^{G}R_{AB}-\frac{{}^G R}{2}G_{AB}=\frac{1}{12}\left[F_{(4)\;ACDE}F_{(4)\;B}^{\phantom{(4)\;B}CDE}-\frac{G_{AB}}{8}F_{(4)\;CDEF}F_{(4)}^{\phantom{(4)\;}CDEF}\right]
\\
& \mathrm{d}\star_{11} F_{(4)}=\frac{1}{2}F_{(4)}\wedge F_{(4)}\,,
\end{align}
\end{subequations}%
where $F_{(4)}=\mathrm{d}C_{(3)}$. Upper case Latin indices are now eleven-dimensional. 
In this case the lower dimensional theories are just Einstein-Hilbert gravity endowed with a negative cosmological constant.
%%%%%%%%%%%%
\section{AdS$_5$}
\label{sec:ads5}
%%%%%%%%%%%%
As a warm up for the more general cases studied below, we begin by taking $A=0$ in Eqs.~(\ref{eq:consi}) and focus on Schwarzschild-AdS$_5$ black holes with a spherical horizon. 
The five-dimensional metric is simply given by (\ref{eq:refanyd}) with $d=4$.
The horizon is the null hypersurface $r=r_+$ where $f$ vanishes linearly. The Hawking temperature, entropy and energy of this black hole spacetime are
\begin{equation}
T_H = \frac{1}{2 \pi  r_+}\left(1+\frac{2 r_+^2}{L^2}\right)\,,\quad S_H = \frac{\pi^2r_+^3}{2 G_5}\quad\text{and}\quad E_H = \frac{3\pi r_+^2}{8 G_5}\left(1+\frac{r_+^2}{L^2}\right)\,,
\end{equation}
respectively. 
It can be readily checked that these quantities satisfy the first law of black hole thermodynamics and that the specific heat of these black holes becomes negative for $r_+/L<1/\sqrt{2}$, divergent at $r_+/L=1/\sqrt{2}$ and positive for $r_+/L>1/\sqrt{2}$. 
Furthermore, the Hawking-Page transition occurs at $r_+=L$, which the black hole solution becoming dominant in the canonical ensemble for $r_+>L$ \cite{Hawking:1982dh}.

To bring our analysis as close as possible to \cite{SSS-1}, we introduce a new coordinate $\rho$ defined by
\begin{equation}
r=\sqrt{r_+^2+(L^2+2 r_+^2)\sinh ^2\rho }\,,
\label{defrho}
\end{equation}
with $\rho\geq0$ corresponding to the right side of the eternal Schwarzschild-AdS$_5$ black hole. 
In these new coordinates the line element (\ref{eq:refanyd}) becomes
\begin{equation}
\mathrm{d}s^2_{5D} = -\frac{(1+2 y_+^2)^2 \cosh ^2\rho}{y_+^2+(1+2 y_+^2) \sinh ^2\rho} \sinh ^2\rho\,\mathrm{d}t^2+L^2\,\mathrm{d}\rho^2+L^2\left[y_+^2+ (1+2 y_+^2)\sinh ^2\rho\right]\mathrm{d}\Omega_3^2\,,
\label{eq:br}
\end{equation}
where we defined $y_+\equiv r_+/L$. 
The above metric is singular at $\rho=0$, since the metric degenerates there. 
This singularity could be avoided if we worked in Euclidean signature and identified the period of Euclidean time appropriately. 
However, here we want to keep the period of the Lorentzian time $t$ as being $T$, so there is a genuine singularity at $\rho=0$. 
There are also curvature singularities, which in Schwarzschild coordinates occur at $r=0$, and are now mapped to the complex plane
\begin{equation}
\rho^{\pm\;j}_{\mathrm{sing}} = \pm\,\i\,\arcsin\left(\frac{y_+}{\sqrt{1+2\,y_+^2}}\right)+\i\,j\,\pi
\end{equation}
with $j\in\mathbb{Z}$.
In the $\rho$ coordinates, the action from \eqref{eq:alld} for the D$3$-branes becomes
\begin{equation}
\i\,I_{\mathrm{D}3}(\rho) = \i\,T \,\pi ^2 L^3 (1+2 y_+^2)\, e^{-\rho }\, \sinh \rho\,\left[y_+^2 \sinh (2 \rho )-(1+y_+^2) \cosh (2 \rho )-y_+^2+1\right]\,.
\label{eq:actiond3rho}
\end{equation}
The analysis of anti-branes follows by symmetry as described in section \ref{sec:bab}.\footnote{
    Explicitly, the weighting in the brane path integral is $\exp(\i\,I_{\mathrm{D}3}(\rho))$ and the weight in the anti-brane path integral is $\exp( - \i\,I_{\mathrm{D}3}(-\rho))$.
}

Searching for saddle points in the complex $\rho$ plane is a tedious exercise, but can nevertheless be done analytically. 
We find three different classes of saddles, which can be written as
\begin{subequations}
\begin{equation}
\rho_{\star}^{I\;j}=-\frac{1}{2}\log X^I+\i\,\left(j-\frac{1}{2}\right)\,\pi
\end{equation}
with $j\in\mathbb{Z}$, $I\in\{1,2,3\}$
and
\begin{align}
&X^1 = \frac{1}{1+2y_+^2}\left\{\cosh\left[\frac{1}{3}\mathrm{arccosh}\,(1+8\,y_+^2+8\,y_+^4)\right]-\frac{1}{2}\right\}\,,
\\
&X^2 = \frac{1}{1+2y_+^2}\left\{\cosh\left[\frac{1}{3}\mathrm{arccosh}\,(1+8\,y_+^2+8\,y_+^4)+\frac{2\pi}{3}\,\i\right]-\frac{1}{2}\right\}\,,
\\
&X^3 = \frac{1}{1+2y_+^2}\left\{\cosh\left[\frac{1}{3}\mathrm{arccosh}\,(1+8\,y_+^2+8\,y_+^4)+\frac{4\pi}{3}\,\i\right]-\frac{1}{2}\right\}\,.
\end{align}%
\end{subequations}
We note that $X^1$ is real, and $X^2$,$X^3$ are complex, with $X^2$ having positive imaginary part and being the complex conjugate of $X^3$. I
t is also straightforward to show that $X^1$, $X^2$ and $X^3$ satisfy
\begin{equation}
X^1+X^2+X^3=-\frac{3}{2(1+2 y_+^2)}\,.
\end{equation}
It is now a simple matter to evaluate the real part of the action (\ref{eq:actiond3rho}) on the corresponding saddles, for which we find
\begin{subequations}
\begin{align}
&\mathrm{Re}\left[\frac{\i}{T L^3}I_{\mathrm{D}3}\left(\rho_{\star}^{1\;j}\right)\right]=0\,,
\label{eq:rho1}
\\
\label{eq:rho2}
&\mathrm{Re}\left[\frac{\i}{T L^3}I_{\mathrm{D}3}\left(\rho_{\star}^{2\;j}\right)\right]=\frac{3 \sqrt{3}\pi ^2}{8} \sinh \left(\frac{2 \chi }{3}\right) \left[\frac{(1+2 y_+^2)^2}{\cosh (2 \chi )}-\cosh \left(\frac{2 \chi }{3}\right)\right]\,,
\\
\label{eq:rho3}
&\mathrm{Re}\left[\frac{\i}{T L^3}I_{\mathrm{D}3}\left(\rho_{\star}^{3\;j}\right)\right]=-\frac{3 \sqrt{3}\pi ^2}{8} \sinh \left(\frac{2 \chi }{3}\right) \left[\frac{(1+2 y_+^2)^2}{\cosh (2 \chi )}-\cosh \left(\frac{2 \chi }{3}\right)\right]\,,
\end{align}
with
\begin{equation}
\chi = \log \left(\sqrt{1+y_+^2}+y_+\right) \,.
\end{equation}%
\end{subequations}
The saddles with the most potential danger are $\rho_{\star}^{2\;j}$, for which  \eqref{eq:rho2} is positive definite; the corresponding quantities are negative definite or identically zero for the other saddles. Note that the action is independent of the $j$ index, as only $\exp(2\rho)$ enters actions and equations.
However, finding a saddle is not enough to show that it actually contributes to the path integral, as discussed in detail in section \ref{sec:saddlesanddescents}.

\begin{figure}[t]
    \centering
    \includegraphics[width=0.7\textwidth]{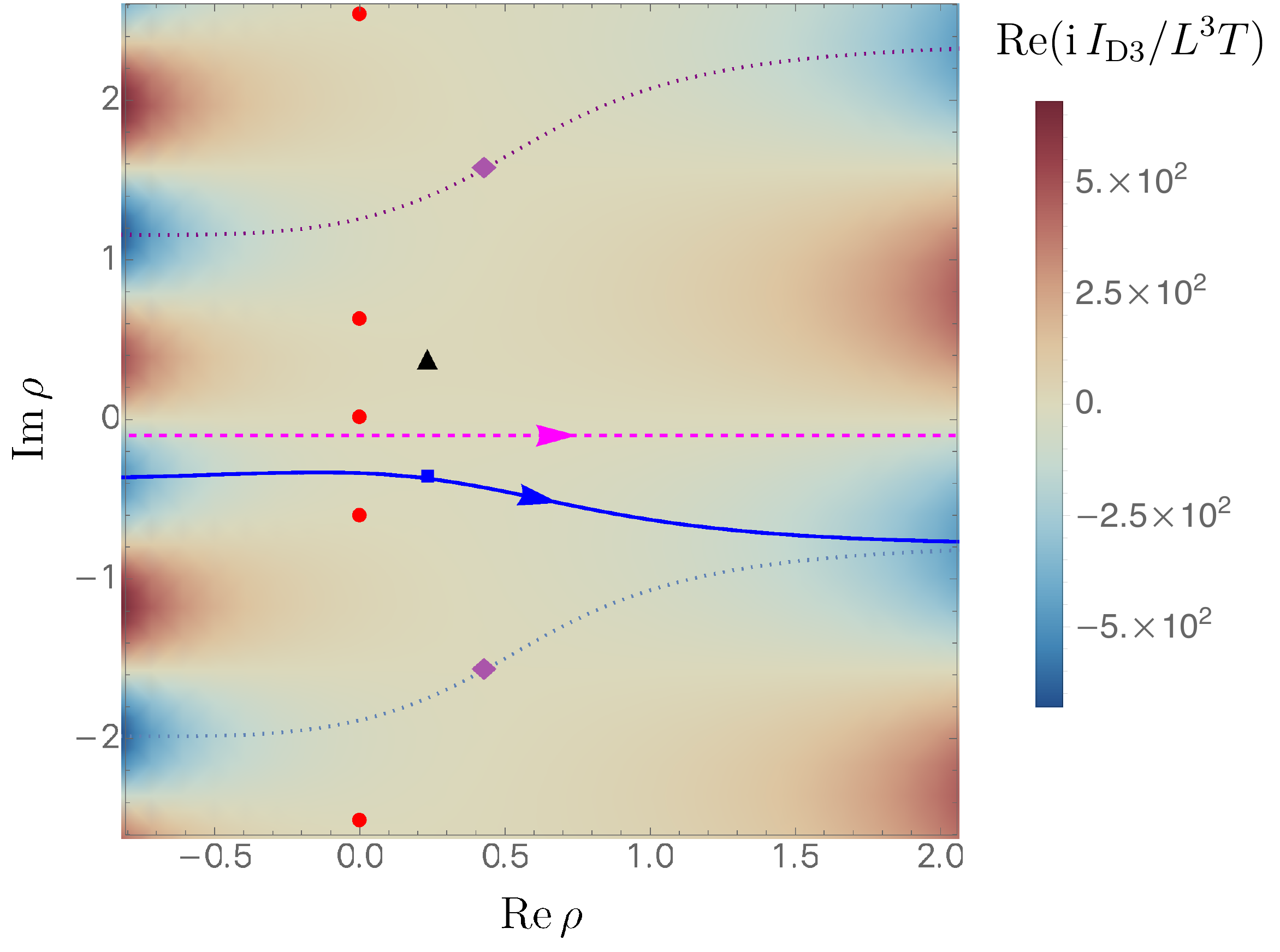}
    \caption{A density plot of $\mathrm{Re}\left(\i I_{D3}/L^3T\right)$ in AdS$_5$ showing saddles and relevant contours for $r_+=L$.
    The red disks are singuarlities, with $\rho=0$ being the orbifold singularity of the double cone at the black hole bifurcation surface.
    The magenta dashed line is slightly below the real axis, and we take it to be the defining contour for our path integral.
    The blue square corresponds to $\rho^{3\;j}_\star$, and the blue line through it is the steepest descent contour.
    The defining contour can be deformed to this, so we find a contribution to the integral from this saddle.  But \eqref{eq:rho3} has negative real part, so this contribution is suppressed; it does not induce a brane nucleation instability.
    The black triangle corresponds to $\rho^{2\;j}_\star$, and the purple diamonds correspond to $\rho^{1\;j}_\star$.
    The steepest descent path through the purple saddle is also shown, but any deformation to the defining contour is obstructed both by regions of large brane amplitude (shaded in red) and the black hole singularity (associated with the red disk below the real axis). The constant phase contour (not shown) that descends from the black triangle in fact also passes through the blue dot, near which it is the contour of steepest ascent; the amplitude then grows without bound as one continues further along this contour. }
    \label{fig:sub}
\end{figure}

Our initial contour $C$ is given by
\begin{equation}
C=\{\rho \in \mathbb{C}:\rho = x-\i\,\varepsilon\text{ with }x\in\mathbb{R}\}\,,
\label{rhocontourequation}
\end{equation}
where $\varepsilon>0$ is taken to be small. 
This contour is chosen in such a way that $e^{\i I_{\mathrm{D}3}}$ decays as $|x|\to+\infty$. 
Furthermore, since $\varepsilon>0$ we avoid the singularity at $\rho=0$.
The initial contour is the dashed magenta line in figure \ref{fig:sub}, it runs parallel to the real $\rho$ axis, and slightly below it.

We now proceed to numerically find the paths of steepest descent associated with each of our saddles.
We show these saddles and a density plot of $\mathrm{Re}\left(\i I_{\mathrm{D}3}(\rho)\right)$ in figure \ref{fig:sub}, taking $r_+=L$.
The red disks indicate the location of the singularities, the singularity at $\rho=0$ is the orbifold singularity at the bifurcation point.
The purple diamonds are saddles of type $\rho^{1\;j}_\star$,
the black triangles are saddles of type $\rho^{2\;j}_\star$,
and the blue squares are saddles of type $\rho^{3\;j}_\star$.
Also shown are the paths of steepest descent through $\rho^{1\;j}_\star$ (dotted purple line) and through $\rho^{3\;j}_\star$ (solid blue line).

We find that we can smoothly deform the original contour to the steepest descent contour through the blue square (the solid blue line), while keeping the contribution from the asymptotic ends suppressed.
This is the saddle we called $\rho^{3\;j}_\star$ (with $j=0$), and on this saddle, we have $\text{Re}(\i I_{D3}) < 0$.\footnote{The $\mathbb{Z}_2$ symmetry discussed in section \ref{sec:bab} implies that the relevant saddle for the anti-brane will be at $-(\rho^{3\;j}_\star)^*$.
Also, as noted in section \ref{sec:bab} and as can be easily checked explicitly in this case, the on-shell value of $\text{Re}\left(\i I_{D3}\right)$ is the same for the brane and anti-brane.}

To double check our calculation, we verified numerically that the integral along the steepest descent path and along our original contour agree to very high accuracy. As described in section \ref{sec:saddlesanddescents}, this means that no other saddle can be more dominant.
But for completeness we include discussion of descent contours through the other saddles in the caption of figure \ref{fig:sub}, making manifest that they cannot be deformed to the defining contour.

Note, however, that the contour of steepest descent through a saddle described by a purple square is also asymptotically suppressed at both ends. 
But the initial contour cannot be deformed to this contour without crossing over one of the red dots in figure \ref{fig:sub} at ${\rm Im} \, \rho \neq 0$, corresponding to the black hole singularity at $r=0$. 
So the purple saddles are not relevant to the desired calculation.

%%%%%%%%%%%%
\section{General dimensions\label{sec:gend}}
%%%%%%%%%%%%
We now establish the absence of brane nucleation instability on the double cone in arbitrary dimensions (as usual, at the strict probe-brane level).
We work in AdS$_{d+1}$ with $d\geq 3$.
Section \ref{sec:ads5} studied the case of $d=4$, but as described in section \ref{sec:coven} the general case is similar and involves branes with $p=d-1$.

We work directly in the $r$-coordinate without attempting to introduce an analogue of the $\rho$-coordinate (\ref{defrho}).
The brane contributes to the path integral a weight given by \eqref{eq:alld}.
Note that $f \approx r^2/L^2$ at large $r$ so that the area and volume terms almost cancel; this is a consequence of working in a supersymmetric model.
The extremization of the brane action yields the equation
\begin{align}
    \frac{1}{2} f' \, r^{d-1} + f r^{d-2} (d-1) = \frac{d\,r^{d-1}}{L}\,\sqrt{f}\, .
    \label{exteqf}
\end{align}
Only the right hand side involves a square root.
The extremization equation can be squared and simplified to a polynomial equation involving four terms.
The saddle point value $r_\star$ satisfies
\begin{equation}
    \frac{r_\star^{2(d-1)} \, d\,(d -2)}{L^2} +r_\star^{2(d-2)} \, (d-1)^2 -2\,r_\star^{d-2} \, d(d-1) \, M+ d^2 M^2 = 0\, .
    \label{saddleeqr}
\end{equation}
We also find a very simple expression for the on-shell action
\begin{align}
    \exp \i I_\star = \exp \left\{ - \frac{\i\,T\,\Omega_{d-1}}{L} \,\left[\frac{d-1}{d} \, L^2\,r_\star^{d-2}-L^2 M +r_+^d\right] \right\} \, .
    \label{onshellsimple}
\end{align}
A saddle that dominates over the configuration with no branes should thus have $\text{Im}\, (r_\star^{d-2}) > 0$.

\begin{figure}[t]
    \centering
    \includegraphics[width=0.7\textwidth]{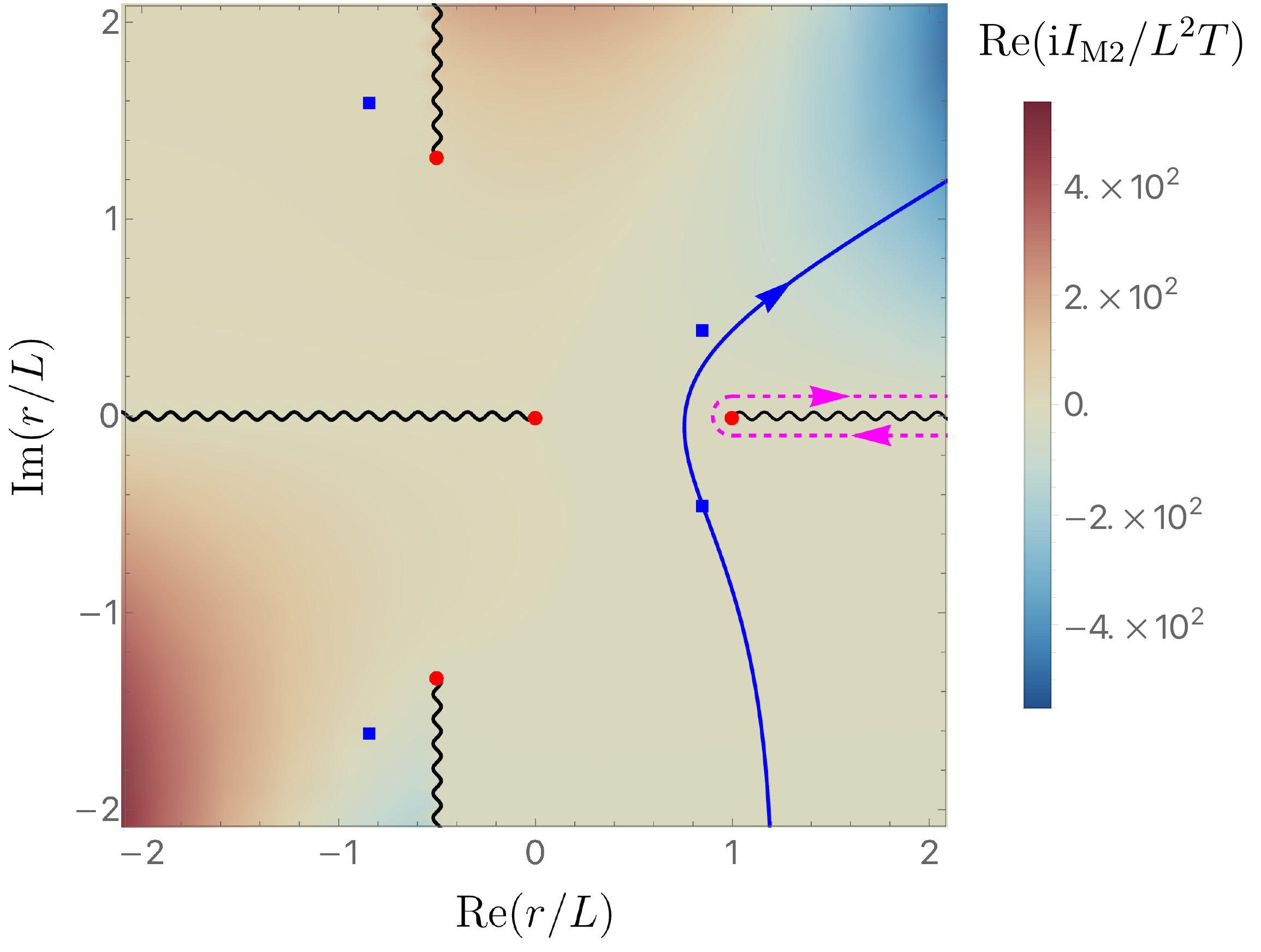}
    \caption{A density plot of $\mathrm{Re}\left(\i I_{\mathrm{M}2}\right)$ in AdS$_4$ showing saddles and relevant contours in the complex $r$ plane. We take $r_+ = L$.
    The red dots are branch points for the action, which now include all singularities of the double cone spacetime (though some branch points also occur at smooth points of the double cone spacetime). The dashed magenta curve is the defining contour for our integral.  The blue dots are saddles, and one of these has a global descent contour (solid blue line) that can be deformed to the dashed magenta curve.  The relevant saddle has  $\mathrm{Re}\left(\i I_{\mathrm{M}2}\right) <0$.  As described in section \ref{sec:saddlesanddescents}, no other saddle can be more dominant (and we have again checked explicitly that the other saddles fail to lie on useful descent contours), so the system is stable to brane nucleation.
    }
    \label{fig:ads4}
\end{figure}

\paragraph{AdS$_4$:}
Figure \ref{fig:ads4}
shows a density plot of density plot of $\mathrm{Re}\left(\i I_{\mathrm{M}2}\right)$ in AdS$_4$ indicating  saddles and relevant contours in the complex $r$ plane for $r_+ = L$. Other values of $r_+$ are similar. As in AdS$_5$, we can identify a saddle on a global descent contour that can be deformed to the defining contour for our integral. And again the relevant saddle has  $\mathrm{Re}\left(\i I_{\mathrm{M}2}\right)<0$.
As described in section \ref{sec:saddlesanddescents}, no other saddle can be more dominant,\footnote{We have also checked explicitly that no other saddle lies on a global descent contour that can be deformed to the defining contour.} so the system is stable to brane nucleation.

\begin{figure}[htb]
    \centering
    \includegraphics[height=0.35\textwidth]{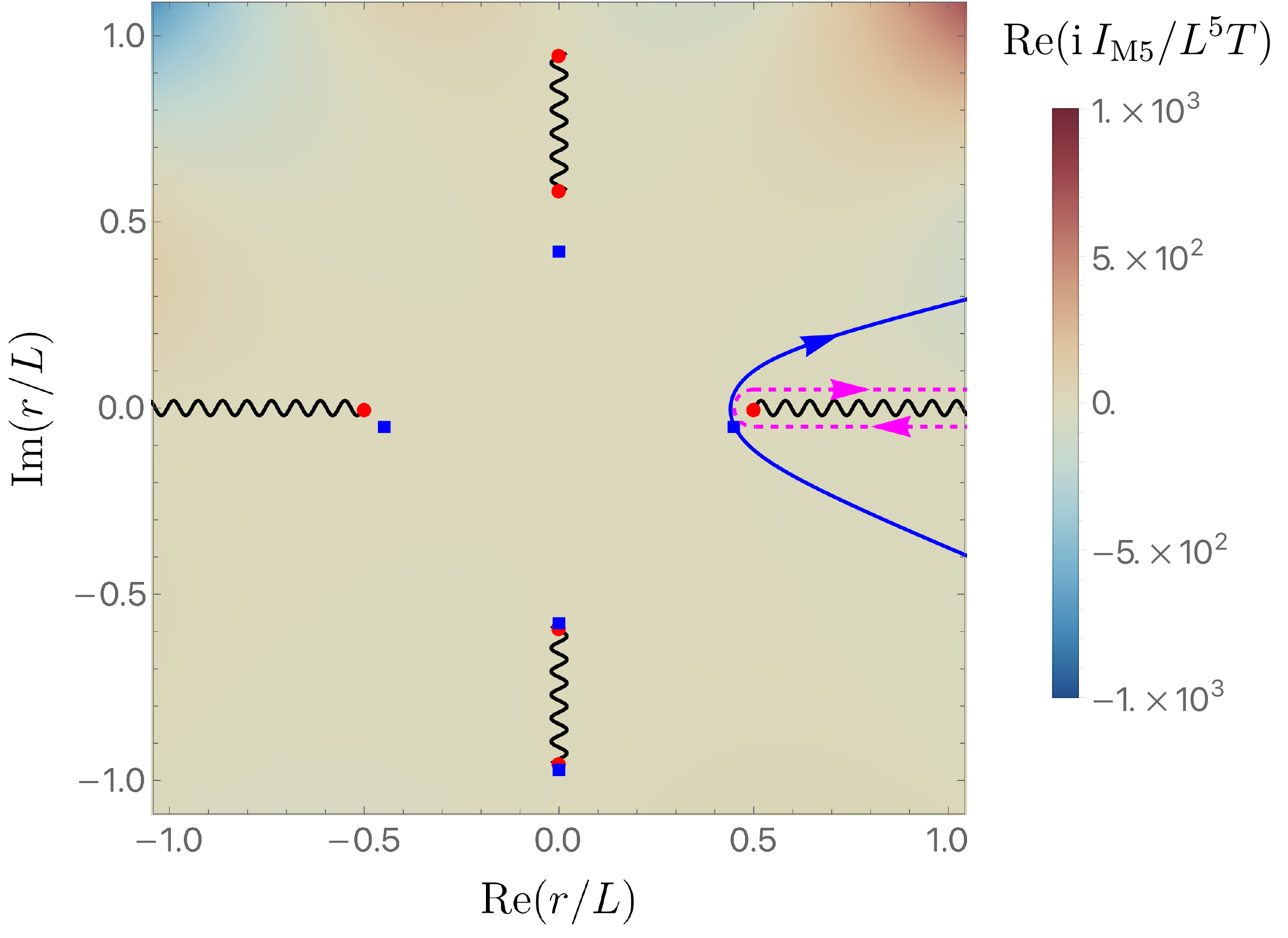}\hspace{0.5cm}\includegraphics[height=0.35\textwidth]{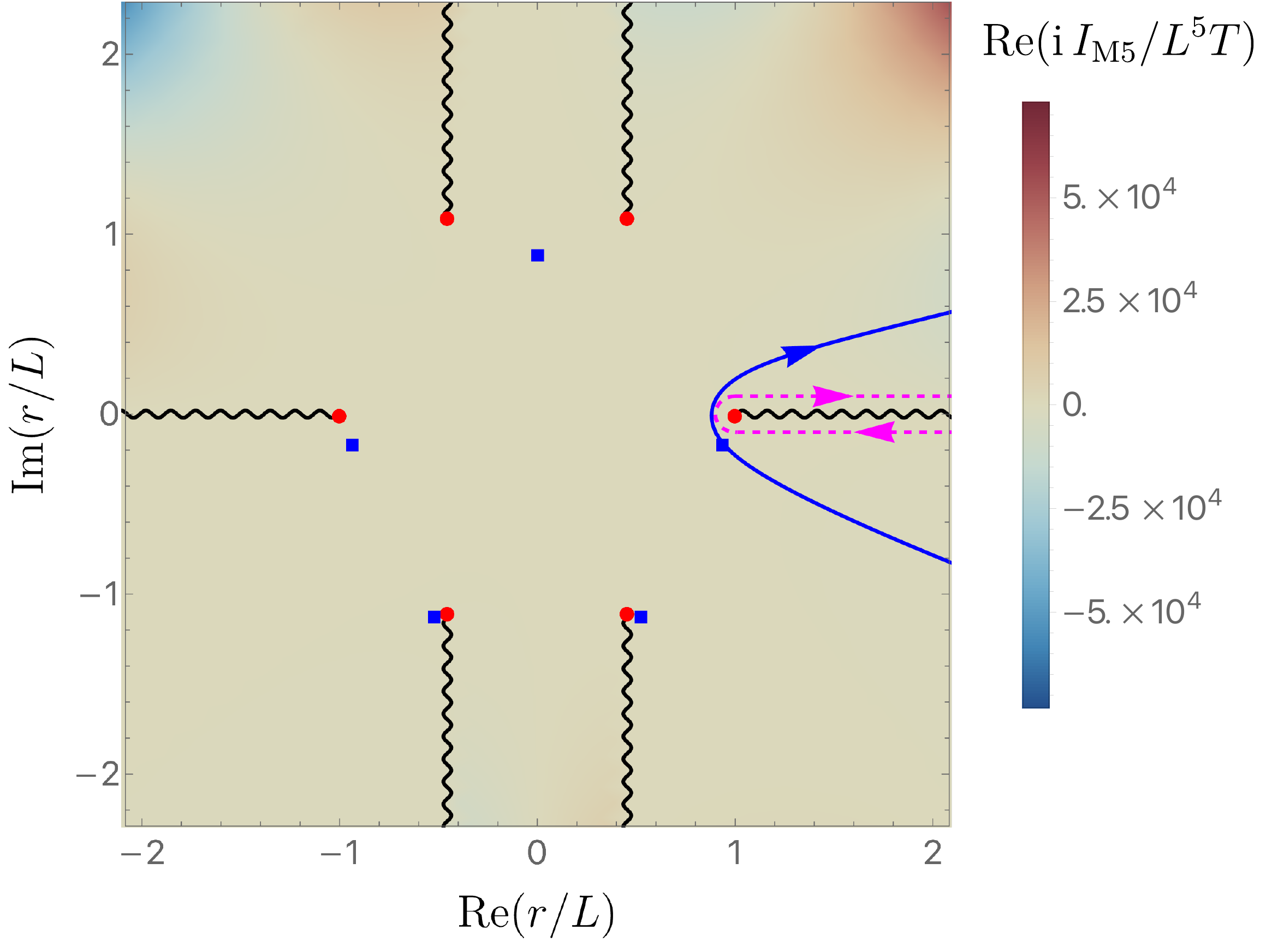}
    \caption{A density plot of $\mathrm{Re}\left(\i I_{\mathrm{M}5}\right)$ in AdS$_7$ showing saddles and relevant contours in the complex $r$ plane.  We use the same conventions as in figure \ref{fig:ads4}, with the steepest descent contour that can be deformed to the defining contour shown in solid blue. Again, its saddle has $\mathrm{Re}\left(\i I_{\mathrm{M}5}\right)<0$ and indicates stability with respect to brane nucleation. The details of the branch point structure change at $r_+/L = \sqrt{3}$. The left panel was generated with $r_+=L/2$, whereas on the right panel we have $r_+=L$. }
    \label{fig:ads7}
\end{figure}

\paragraph{Higher dimensions:}
Increasing $d$ increases the degree of the polynomial equation to be solved for saddles.  
We thus find an increasing number of branch points in the $r$ plane located at roots of $f$. 
Typical cases for AdS$_7$ are shown in figure \ref{fig:ads7}.  
The most relevant structures are similar to that seen in AdS$_4$, especially near the the line (blue curve) of steepest descent that can be deformed to the defining contour. 
However, the detailed structure of branch points and branch cuts depends on $r_+$ as shown in the figures.
As above, the relevant saddle has $\mathrm{Re}\left(\i I_{\mathrm{M}7}\right)<0$ indicating stability with respect to brane nucleation.  
We have checked that these statements remain true for all $3 \le d \le 20$.
This includes many cases that cannot correspond to microscopic supersymmetric theories, but where our universal form can be studied nonetheless.

\paragraph{Large $d$ limit:}
The above numerical results suggest that the double cone is stable to brane nucleation for all $d \ge 3$.
An analytic treatment also becomes possible in the large $d$ limit, where we take the limit $d\to \infty$ while keeping $r_+$ fixed.
We make the ansatz\footnote{We determine all coefficients to $i=4$, when the error in Eq.~(\ref{eq:ansatzrstar}) compared with the exact results in $d=6$ where well under the one percent level.}
\begin{equation}
r_\star=r_+\,\sum_{i=0}^{\infty}\frac{\alpha_{(i)}}{d^i}\,,
\label{eq:ansatzrstar}
\end{equation}
and determine the coefficients $\alpha_{(i)}$ in an expansion of Eq.~(\ref{saddleeqr}) at large $d$. 
We find that $\alpha_{(0)}=1$, which means that as $d\to+\infty$, $r_{*}$ approaches $r_+$. The next to leading order coefficient reads
\begin{equation}
4 e^{2 \alpha_{(1)} }-4 e^{\alpha_{(1)} }+y_+^2+1=0\,,
\end{equation}
where recall that we defined $r_+\equiv y_+\,L$. 
This equation can be readily solved to give
\begin{equation}
\alpha^{j}_{(1)\;\pm} = 2\,\pi\,j\,\i+\log\left[\frac{1}{2}(1\pm\i y_+)\right]\,,
\end{equation}
with $j\in \mathbb{Z}$ and we defined the $\log$ so that $\log z = \log |z|+\i\,\arg z$ with $-\pi< \arg z< \pi$. 
Keeping the root closest to the real axis and with a negative imaginary part (which we expect to be the relevant one, though this is nontrivial to check in detail), we find
\begin{equation}
r_\star = r_+\left\{1+\frac{1}{d}\log\left[\frac{1}{2}(1-\i y_+)\right]+\mathcal{O}(d^{-2})\right\}\,.
\label{saddlerstard}
\end{equation}

As $d \rightarrow \infty$ this value approaches the horizon $r=r_+$ from the lower half plane (i.e.,  with a small imaginary part).
Note also that (\ref{saddlerstard}) gives us the value of the on-shell action (\ref{onshellsimple}) in the large $d$ limit, for which we find
\begin{equation}
\exp \i I_\star = \exp\left\{-\frac{T\,r_+^{d-1}\,\Omega _{d-1}}{2}\,(1+\i\,y_+)\left[1+\mathcal{O}(d^{-1})\right]\right\},
\end{equation}
and we see that the brane action is suppressed.  
This provides a further check on the claim that any $d \ge 3$ is qualitatively similar to the $d=3$ case shown in figure \ref{fig:ads4}.

We thus find the Schwarzschild double cone to be stable to brane nucleation at the strict probe-brane level in any AdS$_{d+1}$ with $d\geq 3$.

%%%%%%%%%%%%
\section{\label{sec:CMP}The Cveti\v{c}-L\"u-Pope black holes}
%%%%%%%%%%%%
Here we study a family of charged Kerr-de Sitter black holes in five dimensions with two equal angular momenta whose line element and gauge field read
\begin{subequations}
\begin{align}
&\mathrm{d}s^2 = -\frac{f}{h}\mathrm{d}t^2+\frac{\mathrm{d}r^2}{f}+\frac{r^2}{4}(\sigma_1^2+\sigma_2^2)+\frac{r^2}{4}h\left(\sigma_3-W\, \mathrm{d}t\right)^2\,,
\label{eq:5Dg}
\\
& A = \frac{\sqrt{3}\tilde{Q}}{r^2}\left(\mathrm{d}t-\frac{\tilde{J}}{2}\sigma_3 \right)\,,
\end{align}
\label{eq:5D}%
\end{subequations}
where $\sigma_1, \sigma_2, \sigma_3$ are the usual left-invariant 1-forms of $S^3$ with
\begin{subequations}
\begin{align}
&\sigma_1 =\cos\psi\sin\theta~\mathrm{d}\phi-\sin\psi~\mathrm{d}\theta \,,
\label{eq:sigma1}
\\
&\sigma_2=\sin\psi\sin\theta~\mathrm{d}\phi+\cos\psi~\mathrm{d}\theta\,,
\label{eq:sigma2}
\\
&\sigma_3=\mathrm{d}\psi+\cos\theta~\mathrm{d}\phi\,,
\end{align}
\label{eq:sigma3}%
\end{subequations}
and
\begin{subequations}
\begin{align}
    &f = \frac{r^2}{L^2}+1-\frac{2 \tilde{M}}{r^2}\left(1-\chi\right)+\frac{\tilde{Q}^2}{r^4}\left(1-\frac{\tilde{J}^2}{L^2}+\frac{2\tilde{M} L^2 \chi}{\tilde{Q}^2}\right)\,,
    \\
    & W = \frac{2\tilde{J}}{r^2 h}\left(\frac{2 \tilde{M}+\tilde{Q}}{r^2}-\frac{\tilde{Q}^2}{r^4}\right)\,,
    \\
    & h = 1-\frac{\tilde{J}^2 \tilde{Q}^2}{r^6}+\frac{2 \tilde{J}^2 (\tilde{M}+\tilde{Q})}{r^4}\,,
\end{align}
\end{subequations}
where $L^2\chi \equiv \tilde{J}^2(1+\tilde{Q}/\tilde{M})$.
The constants $\tilde{M}$, $\tilde{Q}$ and $\tilde{J}$ parameterize the energy $E_H$, electric charge $Q$ and angular momentum $J$ as
\begin{subequations}
\begin{align}
&E_H = \frac{3 \tilde{M}\pi}{4 G_5}\left(1+\frac{\chi}{3}\right)\,,
\\
& Q = \frac{\sqrt{3} L \pi \tilde{Q}}{4 G_5}\,,
\\
& J = \frac{\tilde{J} \pi}{4 G_5}(2\tilde{M}+\tilde{Q})\,.
\end{align}
\end{subequations}

The black hole event horizon is the null hypersurface $r=r_+$, with $r_+$ being the largest real positive root of $f(r)$. 
In the static limit, where $\tilde{J}=0$, this solution reduces to the familiar Reissner-Nordstr\"om black hole with AdS asymptotics, whereas in the uncharged limit, where $\tilde{Q}=0$, it yields the equal angular momenta Kerr-AdS black hole first found in \cite{Hawking:1998kw} and subsquentely in \cite{Gibbons:2004uw}. 
Perhaps more importantly, if we set
\begin{subequations}
\begin{align}
&\tilde{M}=\tilde{M}_{\mathrm{BPS}} \equiv r_+^2 \left(1+\frac{3 r_+^2}{2 L^2}+\frac{r_+^4}{2 L^4}\right)
\\
&\tilde{Q}=\tilde{Q}_{\mathrm{BPS}} \equiv r_+^2 \left(1+\frac{r_+^2}{2 L^2}\right)\,,
\\
&\tilde{J}=\tilde{J}_{\mathrm{BPS}}\equiv \frac{L r_+^2}{r_+^2+2 L^2}\, ,
\end{align}%
\end{subequations}
the solution becomes the one-parameter family of supersymmetric black holes found in \cite{Gutowski:2004yv,Gutowski:2004ez}.
Although the parameters $\tilde{M}$, $\tilde{Q}$, $\tilde{J}$ allow us to easily write all metric functions, it turns out to be more beneficial to write $f$ in terms of $r_+$ and the radius of the Cauchy horizon $r_-\leq r_+$. 
Using the fact that $f$ vanishes at the Cauchy horizon and at the black hole event horizon, we can invert the relation between $\tilde{M}$ and $\tilde{Q}$ as a funtion of $r_+$ and $r_-$. 
In terms of $r_+$ and $r_-$ the function $f(r)$ simplifies considerably, namely
\begin{equation}
f(r)=\frac{(r^2-r_+^2)(r^2-r_-^2)(r^2+L^2+r_-^2+r_+^2)}{L^2 r^4}\,.
\label{eq:frc}
\end{equation}

\begin{figure}[htb]
    \centering
    \includegraphics[width=0.7\textwidth]{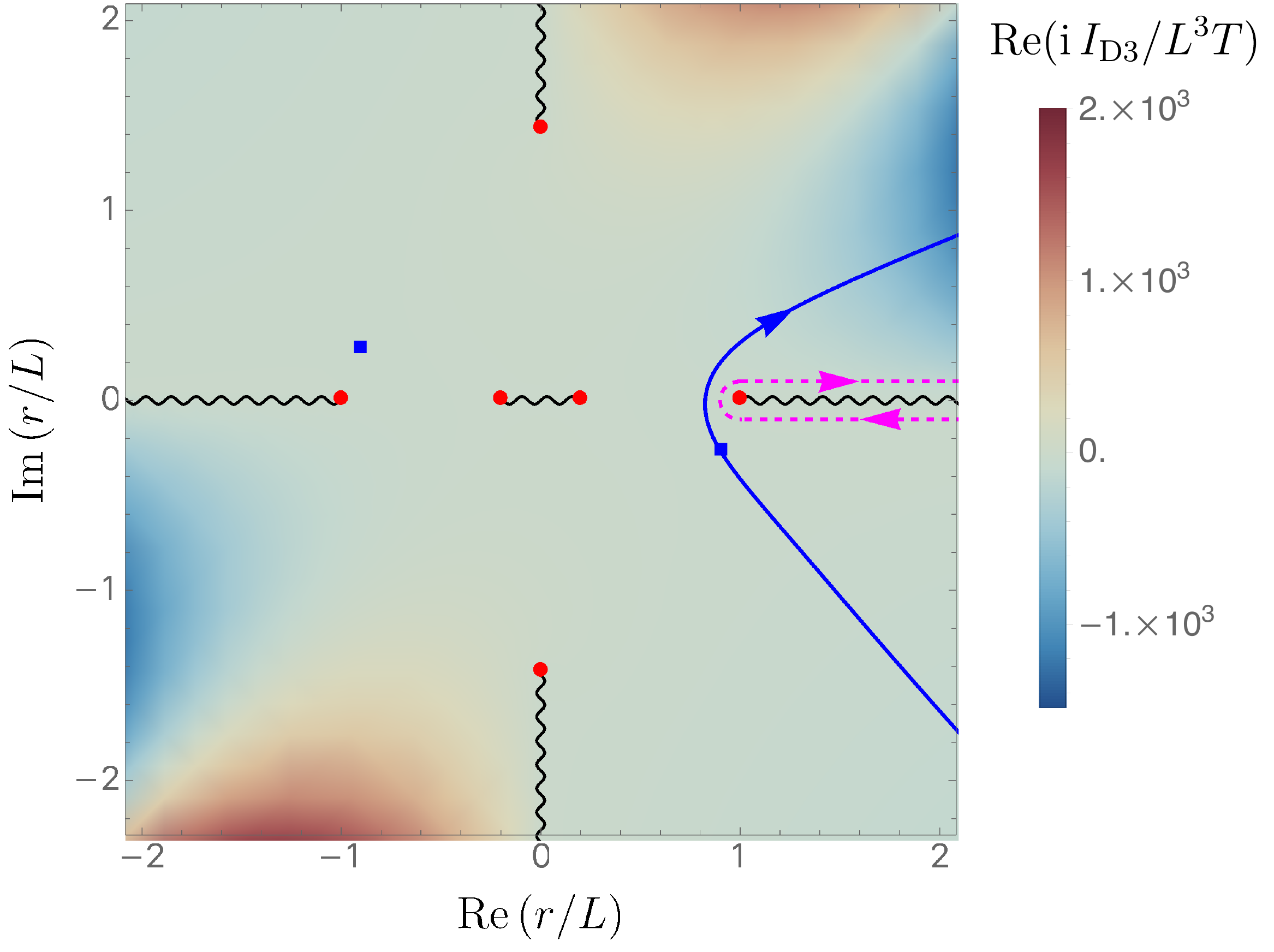}
    \caption{A density plot of $\mathrm{Re}\left(\i I_{\mathrm{D}3}\right)$ for a Cveti\v{c}-L\"u-Pope black hole with $r_+=L$ and $r_-= L/5$.  We use the same conventions as in figure \ref{fig:ads4}, with the (unique!) steepest descent contour that can be deformed to the defining contour shown in solid blue. Again, its saddle (which we call $r_!$) has $\mathrm{Re}\left(\i I_{\mathrm{D}3}\right)<0$.  So as described in section \ref{sec:saddlesanddescents}, no other saddle can be more dominant and our results indicate stability with respect to brane nucleation.}
    \label{fig:subcharged}
\end{figure}

Using our embedding formula (\ref{eq:consi}) together with the D$3$-brane action (\ref{eq:susyborn}) gives an action for D$3$-brane probes that is remarkably similar to (\ref{eq:alld}) with $d=4$, namely
\begin{equation}
\i I_{\mathrm{D}3}(r)=\frac{2 \i \pi ^2 T}{L} \left[r^4-r_+^4-L\,r^3\, \sqrt{f(r)}\right],
\end{equation}
but now with $f(r)$ given as in (\ref{eq:frc}). We are thus left with
\begin{equation}
\i I_{\mathrm{D}3}(r)=\frac{2 \i \pi ^2 T}{L} \left[r^4-r_+^4-r\, \sqrt{(r^2-r_+^2)(r^2-r_-^2)(r^2+L^2+r_-^2+r_+^2)}\right]\,.
\label{eq:actionrot}
\end{equation}
There are six branch points in the complex $r$ plane located at
\begin{equation}
r_{\mathrm{branch}}=\left\{\pm r_+,\pm r_-,\pm\i(L^2+r_-^2+r_+^2)\right\}\,,
\end{equation}
and we choose the branch cuts as in Fig \ref{fig:subcharged}.

It is no longer possible to find the saddle points of (\ref{eq:actionrot}) in closed form because the resulting equations for determining $r_{\star}$ turn out to be a quintic polynomial in $r^2$.  
However they can be found numerically.

Results for $r_+=L$ and $r_-=L/5$ are described in figure \ref{fig:subcharged}.  
For this case we can again find a steepest descent contour that can be deformed to our defining contour, and as above the relevant saddle has $\mathrm{Re}\left(\i I_{D3}\right) <0$.  
So the system is stable to brane nucleation.  
We will refer to this saddle as  $r_{!}$ below, with the notation indicating that it is the only saddle relevant to the stationary phase approximation.

Other values give similar results when either $r_+$ or $r_-$ are small.
However, even our subleading saddle-point contribution disappears when both $r_+$ and $r_-$ are sufficiently large.  
This can be shown analytically by investigating the discriminant of the above-mentioned quintic.

We will not present here a detailed analysis of the discriminant as it requires a rather tedious calculation. 
However, one can prove that for a given value of $r_+> L$, a value of $r_-(r_+)$ exists such that $r_!$ becomes real, signaling a phase transition. 
At the transition, the steepest descent contour through $r_!$ passes through the branch point.  
And after the transition, the descent contour crosses through the cut and can no longer be deformed to the defining contour.  
So then even the contribution of this subdominant saddle disappears.\footnote{
    When this happens, we can no longer argue that other saddles fail to contribute just by studying the contour for $r_!$.  But a direct check of the descent contours for other saddles again shows that they fail to contribute.
}

Furthermore, one can determine $r_-(r_+)$ explicitly as a particular root of a quartic polynomial. 
For $r_+ > L$ this $r_-(r_+)$ becomes less than $r_+$, so the phenomenon becomes relevant to physically allowed black holes.
The case $r_+=L$ is special, as $r_-(r_+) =r_+$, so the saddle merely merges with the branch point $r=r_+$ at extremality ($r_-=L$) but is otherwise complex. 
For $r_+<L$ and $r_-<r_+$, we find that $r_!$ is always complex.

We have also studied the  uplift of the standard Reissner-Nordstr\"om black hole to eleven-dimensional supergravity using the results in \cite{Chamblin:1999tk}. The conclusions are identical to those for Cveti\v{c}-L\"u-Pope black holes described above.

%%%%%%%%%%%%
\section{Discussion} \label{sec:disc}
%%%%%%%%%%%%

Our work above studied the possibility of brane nucleation in the double cone spacetime.
As described in section \ref{sec:over}, at the strict probe-brane level this reduces to finding complex saddle points for the probe-brane action and studying their action.
We studied general microscopic models in M-theory or in string-theory where the double cone has vanishing dilaton and where the vacuum is AdS$_d\times X$ for some compact $X$.
Schwarzschild double cones were studied in sections \ref{sec:ads5} and \ref{sec:gend}, but more general charged and rotating models were also studied in section \ref{sec:CMP}.

In all cases with positively-curved boundaries we find that saddles either yield suppression relative to the original double cone without branes, or else the saddle fails to lie on a global descent contour that can be deformed to the original contour of integration.
This argues that double cones in such theories are stable to brane nucleation, and thus that they dominate the semiclassical bulk path integral for the boundary spectral form factor in the ramp regime $t_{\text{ramp}} \ll T \ll e^{S}$.

Cases with flat (toroidal) boundaries are studied in the appendix.  
While there are no strict instabilities, the brane action remains bounded at infinity.  
As a result, the integral over brane configurations makes the path integral diverge and the double cone does {\it not} dominate the SFF. 
As is also briefly discussed in the appendix, this nevertheless consistent with expectations from the dual field theory (which in this case suffers from an IR divergence in its density of states).

As described in section \ref{sec:coven}, our study was limited in three ways.
First, we considered only brane configurations that preserved the Killing symmetries of the asymptotically AdS$_{d+1}$ factor.
While this restriction is natural, and while it may in fact be that all saddles are of this form, it would be useful for future work to fill in this gap.
Second, we did not investigate the possibility of field-theoretic negative modes.
We mention that \cite{CJnew} will discuss such issues for related constrained instantons, but it would be interesting to probe the issue directly for our complex saddles.

However, the third limitation may be the most interesting.  
This was the fact that we have thus far neglected brane back-reaction, and that -- as discussed in section \ref{sec:bab} -- such back-reaction raises complications for the microcanonical boundary conditions associated with the SFF.  
Recall that considering a brane/anti-brane pair means that we add no net charge and thus preserves the fixed-charge boundary conditions of the double cone background, but that the relevant pairs contribute a net imaginary energy $2\i\, {\rm Im} E_\star$ associated with the Killing field $\partial_t$.  
In perturbation theory this Killing energy becomes related to the difference in ADM energies on the two boundaries, so including back-reaction would appear to violate the boundary conditions that fixed the energy of the original double cone to be the same on both ends.

In many contexts, one could compensate for the addition of Killing energy $2\i \, {\rm Im} E_\star$  by adjusting parameters in the background.  
But the on-shell double cone exists {\it only} when the ADM energies at the two ends agree (so that the total Killing energy defined by $\partial_t$ vanishes).  
So this option is not available.

At the saddle point level one may nevertheless declare victory.  
Since no backgrounds exist that can compensate for the addition of Killing energy $2\i \, {\rm Im} E_\star$, the perturbative addition of such branes cannot lead to new saddles that might dominate over the original double cone.  
Perturbation theory for saddles thus gives no reason to doubt the dominance of the original double cone. 
But this approach is unsatisfying, as it leaves us with no way to analyze the actual physics associated with the addition of such branes.
Let us thus take another look at this issue.

Recall \cite{SSS-1,Marolf:2018ldl} that a microcanonical ensemble can be represented as an integral over canonical ensembles.  
In particular, the microcanonical SFF with energy $E$ and energy-width $\Delta E$ can be written
\begin{equation}
\int d\beta_L d\beta_R \ \langle Z(\beta_L - \i T) Z(\beta_R + \i T)\rangle \  \exp\left((\beta_L +\beta_R)E +(\beta_L^2 + \beta_R^2)\Delta E^2  \right).
\end{equation}
While there is no bulk saddle for $\langle Z(\beta_L - iT) Z(\beta_R + iT)\rangle$, in analogy with either the lower-dimensional results of \cite{SSS-1} or the constrained-instanton results of \cite{Cotler:2020lxj} one may expect to be able to write 
\begin{equation}
\langle Z(\beta_L - \i T) Z(\beta_R + \i T)\rangle = \int db \ e^{-I_{\text{grav}}(\beta_L-\i T , \beta_R+\i T, b)} + \dots
\end{equation}
for some Euclidean gravitational effective action $I_{\text{grav}}(\beta_L -\i T, \beta_R +\i T, b)$ associated with off-shell double cone contributions and where $+\dots$ indicates additional contributions, such as from double cones with additional branes.  
In particular, including a term with a single brane/anti-brane pair would yield
\begin{align}
\label{eq:babbackreact}
\langle Z(\beta_L -\i T) &Z(\beta_R + \i T)\rangle = \int db \ e^{-I_{grav}(\beta_L -\i T, \beta_R+\i T, b)} \notag \\ 
&\times \left( 1  + \int d\rho_{b} d\rho_{\overline{b}} \ e^{-\left[ I_{\text{brane}}(\rho_{b}; \beta_L-\i T, \beta_R+\i T, b) + I_{\text{anti-brane}}(\rho_{\overline{b}}; \beta_L-\i T, \beta_R+\i T, b) \right]} + \dots\right). 
\end{align}

A full analysis would then consist of performing the integrals in \eqref{eq:babbackreact} and comparing the results for the terms with and without branes.  
It would be interesting to return to this task in future work.
However, given that the ramp is relevant for large values of $T$ (so long as $T \ll e^S$), a full analysis may require a great deal of control over the effective action $I_{\text{grav}}(\beta_L -\i T, \beta_R +\i T, b)$ summarizing off-shell contributions.  
In particular, suppose that $I_{\text{grav}}$ is evaluated by performing some of the integrations in our path integral in the stationary phase approximation.  
In simple contexts like those studied in \cite{Cotler:2020lxj}, the result becomes independent of $T$ due to cancellations between the left and right sides of the double cone.  
But general off-shell contributions may nevertheless depend on $T$.  
Though they will come with a small coefficient which vanishes in the semiclassical limit, such terms may nevertheless become relevant at exponentially large times.  
We are thus unable to rule out novel behavior at large $T$, including the potential for new instabilities.

Of course, even if stability of the double cone were to be fully confirmed, the precise implications for AdS/CFT would remain a subject of debate.
The double cone contribution to the SFF gives a precisely-linear ramp and violates boundary factorization.
So unless further UV ingredients in the bulk can provide large corrections, this would then require the bulk theory to be dual to an ensemble of CFTs; see associated comments in section \ref{sec:intro}.
On the other hand, if e.g. bulk string field theory could provide such corrections, then one would expect them to introduce detailed structure in the ramp regime but to nevertheless maintain the simple qualitative features associated with the double cone at some coarse-grained level.
In particular, to be consistent with expectations from quantum chaos, the double cone would still need to dominate calculations of the spectral form factor when one averages the time parameter $T$ over windows that are small on macroscopic scales but large on microscopic scales; see e.g. \cite{SaadTalk}. 
In either case, our results indicate that the double cone will remain a robust part of the description of the spectral form factor in UV-complete higher-dimensional models.

%%%%%%%%%%%%
\subsection*{Acknowledgments}
%%%%%%%%%%%%
We would like to thank Andreas Blommaert, Jorrit Kruthoff, Henry Maxfield, Steve Shenker and Douglas Stanford for helpful discussions.  
DM would particularly like to thank Brianna Grado-White and Claire Keckley for many conversations about the (complex) geometry of the double cone.
RM is supported in part by Simons Investigator Award \#620869.
DM was supported by NSF grant PHY-1801805 and funds from the University of California.
JES is partially supported by STFC consolidated grants ST/P000681/1 and ST/T000694/1.
%%%%%%%%%%%%

%%%%%%%%%%%%
\appendix
%%%%%%%%%%%%
\section{\label{app:btz}BTZ and higher-dimensional black holes with torus boundaries}
%%%%%%%%%%%%
In this appendix we consider the case when the boundary has topology $\mathbb{T}^d\equiv (S^1)^d$ and is thus flat. We are particularly interested in the cases with $d=2,3,4,6$.

The $d=2$ case corresponds to a BTZ black hole \cite{Banados:1992wn}, whose uplift to ten-dimensional type IIB supergravity we have not yet discussed. To this end we consider type IIB supergravity with only the ten-dimensional metric ${}^{(10)}g$, Ramond-Ramond three-form $F_{(3)} \equiv \mathrm{d}A_{(2)}$ and dilaton $\phi$. The corresponding equations of motion are
\begin{subequations}
\begin{align}
& {}^{(10)}R_{AB} = \frac{1}{2}{}^{(10)}\nabla_A\phi{}^{(10)}\nabla_B\phi+\frac{e^\phi}{4}\left[F_{(3)\,ACD}F_{(3)\,B}^{\phantom{(3)B}CD}-\frac{{}^{(10)}g_{AB}}{12}F_{(3)\,CDE}F_{(3)}^{\phantom{(3)}\,CDE}\right]\,,
\\
& \mathrm{d}\left(e^{\phi}\star_{10} F_{(3)}\right)=0\,,
\\
&\Box \phi-\frac{e^{\phi}}{12} F_{(3)\,ABC}F_{(3)}^{\phantom{(3)}\,ABC}=0\,,
\end{align}
\label{eq:IIBSUGRA}%
\end{subequations}%
where $\star_{10}$ is the Hodge operation associated with the ten-dimensional metric ${}^{(10)}g$, ${}^{(10)}\nabla$ its associated metric-compatible connection and upper case Latin indices are ten-dimensional.

Consider the following ten-dimensional field configuration
\begin{subequations}
\begin{equation}
\mathrm{d}s^2 = g_{ab}\mathrm{d}x^a\mathrm{d}x^b+L^2 \mathrm{d}\Omega_3^2+\mathrm{d}z_1^2+\mathrm{d}z_2^2+\mathrm{d}z_3^2+\mathrm{d}z_4^2
\end{equation}
\begin{equation}
F_{(3)} = -\frac{2}{L}\mathrm{Vol}_{3}+2\,L^2\,\mathrm{d}^3\Omega_3
\end{equation}
and
\begin{equation}
\phi = 0\, ,
\end{equation}
\label{eqs:full10d}%
\end{subequations}%
where $\mathrm{d}^3\Omega_3$ is the volume form on a round $3-$sphere, $\mathrm{Vol}_{3}$ is the volume form of the three-dimensional metric ${}^{(10)}g$.  Inserting the \emph{Ans\"atze} (\ref{eqs:full10d}) into the $10$-dimensional equations equations of motion (\ref{eq:IIBSUGRA}) yields a set of three-dimensional equations for $g$ which can be derived from the following three-dimensional action
\begin{equation}
S_{3d} = \int_{\mathcal{M}}\mathrm{d}^3 x\sqrt{-g}\left(R+\frac{1}{L^2}\right)\,.
\label{eq:add}
\end{equation}
In particular, the BTZ black hole can be uplifted to a solution of ten-dimensional type IIB supergravity following the formulae above.

Returning to general $d$, all the metrics we wish the investigate take the following form
\begin{subequations}
\begin{equation}
\mathrm{d}s^2=-f(r)\mathrm{d}t^2+\frac{\mathrm{d}r^2}{f(r)}+r^2 \sum_{i=1}^{d-1}\mathrm{d}x_i^2\,,
\label{eq:torus}
\end{equation}
with
\begin{equation}
f(r)=\frac{r^2}{L^2}-\frac{r_+^{d}}{L^2 r^{d-2}}
\end{equation}
\end{subequations}
where $x_i$ are coordinates on $\mathbb{T}^d$ and the black hole event horizon is the null hypersurface $r=r_+$. Before proceeding we introduce a new coordinate $\rho$ defined via
\begin{equation}
r = r_+\,\cosh ^{\frac{2}{d}}\left(\frac{d\,\rho }{2}\right)\,.
\end{equation}
with $\rho=0$ being the birurcating Killing surface. 
In terms of the $\rho$ coordinates the line element (\ref{eq:torus}) becomes
\begin{equation}
\mathrm{d}s^2=-\frac{r_+^2}{L^2} \cosh ^{\frac{4}{d}}\left(\frac{d\,\rho }{2}\right) \tanh ^2\left(\frac{d\,\rho }{2}\right) \mathrm{d}t^2+L^2\mathrm{d}\rho^2+r_+^2 \cosh ^{\frac{4}{d}}\left(\frac{d\,\rho }{2}\right) \sum_{i=1}^{d-1}\mathrm{d}x_i^2\,.
\end{equation}
It is now a simple exercise to compute the on-shell brane action for D$1$-branes ($d=2$), M$2$-branes ($d=3$), D$3$-branes ($d=4$) and M$5$-branes ($d=6$) for which we find
\begin{equation}
\i I_{d}=\frac{\i\,r_+^d\,T\,\mathrm{V}_{\mathbb{T}^{d-1}}}{2 L}\left(e^{-d\,\rho }-1\right)
\end{equation}
where $\mathrm{V}_{\mathbb{T}^{d-1}}$ is the volume of $\mathbb{T}^{d-1}$.

At large positive $\rho$, the action approaches a constant, and thus the brane contribution to the path integral is divergent. 
This is a well-known effect; see e.g. \cite{Maldacena:1998uz,Seiberg:1999xz,Maldacena:2000hw}. 
In particular, for $d=2$ the theories dual to AdS$_3$ that have BPS branes are 2d sigma models with a non-compact target space.
The density of states, being proportional to the volume of the non-compact target space times a continuous function, is thus, strictly speaking, infinite.
Hence, we do not expect a ramp in the spectral form factor.

If we consider higher dimensional AdS black holes with toroidal boundaries, the contribution to the path integral from brane configurations again diverges.  
The interpretation in the dual field theory is similar.

\bibliographystyle{apsrev4-1long}
\bibliography{bibfile}
\end{document}